%% file: main.tex
\documentclass{aa}  

\usepackage{epstopdf}
\usepackage{mathtools}
\usepackage{mathrsfs}
\usepackage{calrsfs}
\usepackage{natbib}
\usepackage[autostyle, english = american]{csquotes}
\MakeOuterQuote{"}
\usepackage{bm}
\usepackage{graphicx}
\usepackage{txfonts}

\usepackage{url}
\usepackage{hyperref}

\newcommand{\V}{{\bm V}}
\newcommand{\G}{{\bm G}}

\newcommand{\J}{{\bm J}}

\hypersetup{
    colorlinks,
    linkcolor={blue},
    citecolor={blue},
    urlcolor={blue}
}

\begin{document} 

\title{Observational Evidence to Support a Dense Ambient Medium Shaping the Jet in 3C 84}

    \author{Jongho Park
           \inst{1,2,3}, 
           Motoki Kino
           \inst{4,5}, 
           Hiroshi Nagai
           \inst{5,6},
           Masanori Nakamura
           \inst{7,3},
           Keiichi Asada
           \inst{3},
           Minchul Kam
           \inst{8},
           \and
           Jeffrey A. Hodgson
           \inst{9}
           }

    \institute{Department of Astronomy and Space Science, Kyung Hee University, 1732, Deogyeong-daero, Giheung-gu, Yongin-si, Gyeonggi-do 17104, Republic of Korea\\
              \email{jparkastro@khu.ac.kr}
        \and
            Korea Astronomy and Space Science Institute, 776 Daedeok-daero, Yuseong-gu, Daejeon 34055, Republic of Korea
        \and
            Institute of Astronomy and Astrophysics, Academia Sinica, P.O. Box 23-141, Taipei 10617, Taiwan
        \and
            Kogakuin University of Technology \& Engineering, Academic Support Center, 2665-1 Nakano-machi, Hachioji, Tokyo 192-0015, Japan
        \and
            National Astronomical Observatory of Japan, Osawa 2-21-1, Mitaka, Tokyo 181-8588, Japan
        \and
            The Graduate University for Advanced Studies, SOKENDAI, Osawa 2-21-1, Mitaka, Tokyo 181-8588, Japan
        \and
            National Institute of Technology, Hachinohe College, 16-1 Uwanotai, Tamonoki, Hachinohe, Aomori 039-1192, Japan
        \and
            Department of Physics and Astronomy, Seoul National University, Gwanak-gu, Seoul 08826, Republic of Korea
        \and
            Department of Physics and Astronomy, Sejong University, 209 Neungdong-ro, Gwangjin-gu, Seoul 05006, Republic of Korea
             }
   \date{Received \today; accepted \today}

  \abstract
  {Highly collimated relativistic jets are a defining feature of certain active galactic nuclei (AGN), yet their formation mechanism remains elusive. Previous observations and theoretical models have proposed that the ambient medium surrounding the jets could exert pressure, playing a crucial role in shaping the jets. However, direct observational confirmation of such a medium has been lacking. In this study, we present very long baseline interferometric (VLBI) observations of 3C 84 (NGC 1275), located at the center of the Perseus Cluster. Through monitoring observations with the Very Long Baseline Array (VLBA) at 43 GHz, a jet knot was detected to have been ejected from the sub-parsec scale core in the late 2010s. Intriguingly, this knot propagated in a direction significantly offset from the parsec-scale jet direction. To delve deeper into the matter, we employ follow-up VLBA 43 GHz observations, tracing the knot's trajectory until the end of 2022. We discovered that the knot abruptly changed its trajectory in the early 2020s, realigning itself with the parsec-scale jet direction. Additionally, we present results from an observation of 3C 84 with the Global VLBI Alliance (GVA) at 22 GHz, conducted near the monitoring period. By jointly analyzing the GVA 22 GHz image with a VLBA 43 GHz image observed about one week apart, we generated a spectral index map, revealing an inverted spectrum region near the edge of the jet where the knot experienced deflection. These findings suggest the presence of a dense, cold ambient medium characterized by an electron density exceeding $\sim10^5\ {\rm cm^{-3}}$, which guides the jet's propagation on parsec scales and significantly contributes to the overall shaping of the jet.
  }

   \titlerunning{Observational Evidence for a Dense Ambient Medium Shaping the Jet}
   \authorrunning{Park et al.}
   
   \maketitle
%
%-------------------------------------------------------------------

\section{Introduction}
\label{sec:introduction}
\input{introduction}

\section{Observations and Data Reduction}
\label{sec:observations}

\input{observations}

\section{Analysis \& Results}
\label{sec:analysis}

\input{analysis}

\section{Discussion}
\label{sec:discussion}
\input{discussion}

\section{Summary}
\label{sec:summary}
\input{summary}

\begin{acknowledgements}
\input{acknowledgements}
\end{acknowledgements}

\bibliographystyle{aa}
\bibliography{main.bib}

\begin{appendix} 
\label{appendix}
\input{appendix}

\end{appendix}

\end{document}

%% file: introduction.tex
It is widely accepted that supermassive black holes reside at the centers of most galaxies \citep[e.g.,][]{KH2013}. Some of these black holes emit vast amounts of radiation due to the immense gravitational forces acting on surrounding matter, resulting in the formation of active galactic nuclei (AGN; e.g., \citealt{Netzer2013}). Among AGN, a fraction exhibit highly collimated relativistic jets \citep[e.g.,][]{Blandford2019}. These jets are believed to be launched through complex interactions involving accreting matter, magnetic fields around the black holes, and the rotational motion of the black holes themselves \citep[e.g.,][]{BZ1977, BP1982}. The impact of these jets extends significantly to the evolution of host galaxies and clusters \citep{Fabian2012}. Despite numerous efforts, understanding the mechanism behind the formation of these highly collimated relativistic jets in AGNs remains a longstanding and intriguing puzzle.

Previous observations of the jet in the nearby radio galaxy M87, carried out with the Very Long Baseline Array (VLBA), have revealed that the jet exhibits a semi-parabolic shape within the Bondi radius, transitioning into a conical shape outside this radius \citep{AN2012, Hada2013, NA2013}. This led to the suggestion that the M87 jet is collimated by the pressure of the ambient medium surrounding it, with its dynamics influenced by the black hole's gravity \citep{Nakamura2018}. Among the natural candidates for this medium are winds, non-relativistic gas outflows that can be naturally produced in hot accretion flows \citep[e.g.,][]{Sadowski2013, YN2014, Yuan2015, Nakamura2018}. In support of this idea, recent general relativistic magnetohydrodynamic (GRMHD) simulations have demonstrated that the observed jet collimation in M87 can indeed be produced by the pressure of winds, highlighting the significance of the ambient medium in shaping AGN jets \citep{Nakamura2018, Chatterjee2019}. Additionally, recent polarimetric multifrequency VLBA observations in M87 have found that the magnitudes of Faraday rotation measure (RM) systematically decrease with increasing distance from the black hole \citep{Park2019a}. The inferred pressure profile of the ambient medium surrounding the jet, based on the RM profile, aligns with models predicting jet collimation due to the ambient medium \citep[e.g.,][]{Komissarov2009, Lyubarsky2009}. Moreover, various AGN have exhibited analogous transitions in jet geometries, as revealed by Very Long Baseline Interferometric (VLBI) observations \citep[e.g.,][]{Tseng2016, Hada2018, Nakahara2018, Kovalev2020, Boccardi2021, Park2021b, Okino2022}. These findings indicate that the influence of the ambient medium may play a critical role in shaping AGN jets universally.

The radio source 3C 84 (Perseus A) is situated within the elliptical galaxy NGC 1275, serving as the central galaxy of the Perseus cluster. Investigating the impact of the ambient medium on AGN jet shaping becomes particularly promising with 3C 84 due to its unique characteristics. Its proximity, with a redshift of $z=0.0176$ \citep{Strauss1992}, create a scale where 1 milliarcsecond (mas) corresponds to 0.36 pc for 3C 84. This scale enables VLBI observations capable of effectively resolving the region responsible for the initial jet formation and the associated ambient medium.

3C 84 exhibits long-term variability in the radio band \citep[e.g.,][]{Paraschos2023}. In the 1970s and 1980s, there were flux density outbursts, followed by two decades of gradual decrease. However, in the 2000s, the flux density started to increase, coinciding with the emergence of a new jet component from the sub-parsec scale core \citep{Nagai2010} and the detection of $\gamma$-ray emission by the Large Array Telescope of the Fermi Gamma-ray Space Telescope in the late 2000s \citep{Abdo2009}. Subsequently, in the late 2010s, a jet knot was detected to have been ejected from the core, based on VLBA monitoring observations at 43 GHz \citep{Punsly2021, Paraschos2022}. Remarkably, this knot propagated in the South-Eastern direction, significantly offset from the parsec-scale jet direction (South). The apparent speed of this knot was measured to be approximately 0.1 times the speed of light. This intriguing phenomenon raises questions about the connection between this peculiarly propagating blob and the highly collimated downstream jet.

The VLBI technique enables the attainment of extremely high angular resolution by employing widely separated telescopes. The angular resolution of the VLBI array is primarily determined by its longest baselines, while the image fidelity is influenced by the $(u,v)$-coverage \citep{Thompson2017}. Therefore, a straightforward approach to enhance the performance of a VLBI array involves incorporating as many baselines as possible to improve both the angular resolution and image fidelity. This concept is naturally achieved by combining all VLBI telescopes worldwide to form a single VLBI array, which is precisely what the Global VLBI Alliance (GVA\footnote{\url{http://gvlbi.evlbi.org}}) embodies. The GVA comprises various VLBI arrays from around the world, including the European VLBI Network (EVN), the VLBA, the East Asia VLBI Network (EAVN; \citealt{Cui2021, EAVN2022}), and the Australian Long Baseline Array (LBA). Although the GVA has been involved in observations \citep[e.g.,][]{Bruni2017, Bruni2021, Vegagarcia2020, Johnson2021, Baczko2022, Kim2023}, most of such observations were part of space-VLBI observations, and the full potential of the GVA has not been extensively highlighted yet.

In this study, we aim to investigate the evolution of the knot ejected in the late 2010s and focus on understanding how this knot, initially propagating in an unconventional direction, can be connected to the highly collimated downstream jet. To accomplish this, we utilize the GVA to obtain a high-resolution spectral index map, which enables us to study the role of the ambient medium in shaping the jet.

The paper is structured as follows: In Section~\ref{sec:observations}, we provide details about the observations and data reduction processes. Section~\ref{sec:analysis} presents the results of our analysis regarding jet kinematics and spectral index distribution. Subsequently, in Section~\ref{sec:discussion}, we discuss the obtained results. Finally, in Section~\ref{sec:summary}, we summarize our findings. For our calculations, we adopt an angular diameter distance of 74.216 Mpc for 3C 84, derived from the following cosmological parameters: $H_0 = 69.6$, $\Omega_m = 0.286$, and $\Omega_{\Lambda} = 0.714$ \citep{Bennett2014}.

%% file: observations.tex
\subsection{VLBA Monitoring Observations at 43 GHz}

We present the results obtained from observing 3C 84 using the VLBA at 43 GHz as part of the VLBA-BU Blazar Monitoring Program and the BEAM-ME program\footnote{\url{https://www.bu.edu/blazars/BEAM-ME.html}} \citep{JM2016, Jorstad2017, Weaver2022}. Our focus is on the data collected between October 2019 and November 2022, aiming to investigate the knot ejected from the sub-parsec scale core during the late 2010s \citep{Punsly2021, Paraschos2022}. To generate images of the 3C 84 jet, we employed the calibrated data and CLEAN models provided by the database.

\subsection{Global VLBI Alliance Observation at 22 GHz}

On November 9, 2022, we conducted an observation of 3C 84 using the Global VLBI Alliance (GVA; \citealt{VLBI2030}) at 22 GHz (Project Code: GP060). Initially, a total of 30 telescopes around the world were scheduled to participate in the observation: 10 VLBA stations (Station Codes: BR, FD, HN, KP, LA, MK, NL, OV, PT, SC), the phased VLA (YY), 10 EVN stations (EF, JB, O6, HH, MC, NT, TR, YS, MH, RO), 5 EAVN stations (KY, KU, KT, UR, T6), and 4 Long Baseline Array (LBA) stations (AT, MP, HO, CD). However, due to weather and technical issues, two VLBA stations (BR, KP), two EVN stations (TR, RO), two EAVN stations (UR, T6), and two LBA stations (HO, CD) were unable to participate in the observation. The data were recorded in both right- and left-hand circular polarizations with two-bit quantization in eight baseband channels, each with a bandwidth of 32 MHz. The recording rate was configured at 2 Gbps, except for the LBA, which was set at 1 Gbps. The correlation was performed at the Joint Institute for VLBI ERIC (JIVE) in the Netherlands.

\begin{figure}[t]
\centering
\includegraphics[width=\linewidth]{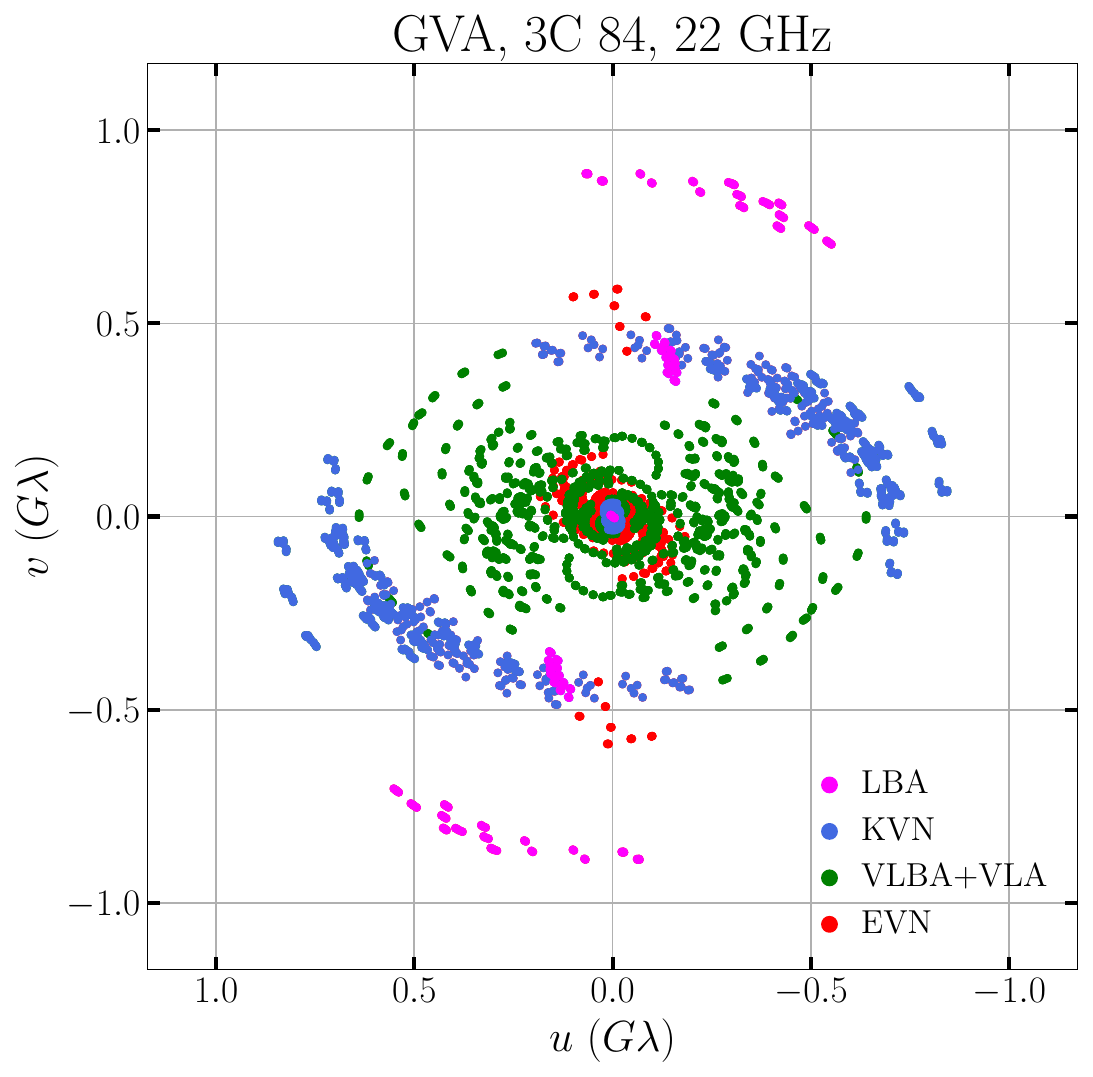}
\caption{The $(u,v)$-coverage of the GVA 22 GHz data is illustrated, highlighting the baselines from LBA, KVN, VLBA+VLA, and EVN in magenta, blue, green, and red colors, respectively. The data has been averaged over the entire bandwidth.}
\label{fig:uvcoverage}
\end{figure}

\begin{figure}[t]
\centering
\includegraphics[width=\linewidth]{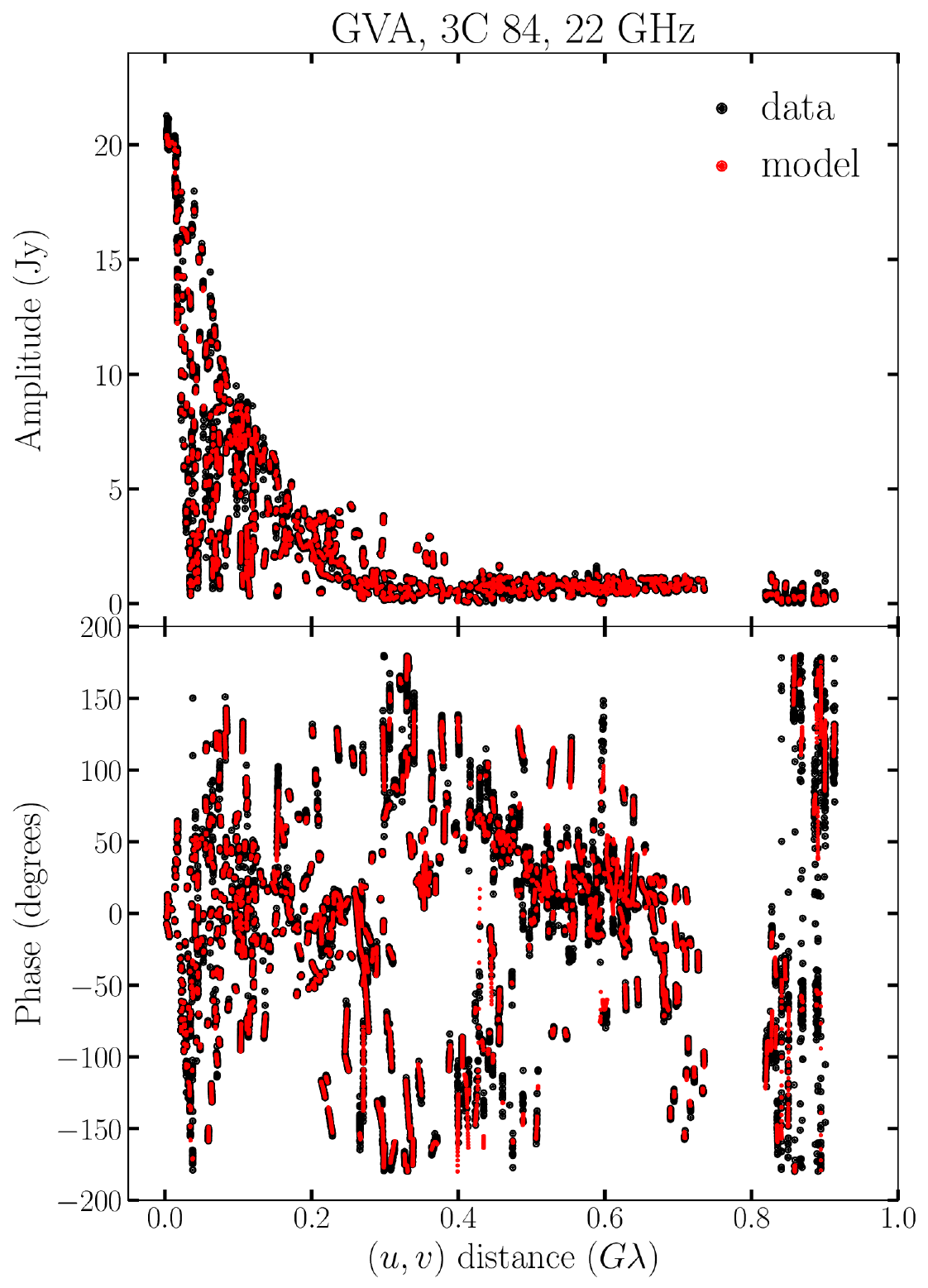}
\caption{The visibility amplitudes (top) and phases (bottom) of the self-calibrated data (black) and CLEAN model (red) are displayed as functions of $(u,v)$-distance in units of $10^9$ wavelengths. The data has been averaged across the entire bandwidth.}
\label{fig:radplot}
\end{figure}

We conducted a standard data reduction using the NRAO's Astronomical Image Processing System (AIPS; \citealt{Greisen2003}), following established procedures described in previous studies \citep{Park2021b}. An important consideration is the heterogeneous nature of the array, composed of multiple widely separated telescopes, which necessitated diligent measures to remove instrumental delay residuals and phase offsets among the baseband channels. Further information regarding the data reduction process and the array's performance will be presented in a separate publication (Park et al., in preparation). Given the high brightness of 3C 84, we employed global fringe fitting \citep{CS1983} with a solution interval of 10 seconds for each baseband channel, successfully detecting fringes across all stations. Subsequently, we performed an iterative process of CLEAN and phase/amplitude self-calibration using the Caltech Difmap package \citep{Shepherd1997}, ultimately generating a source image. We present the $(u,v)$-coverage of the data in Figure~\ref{fig:uvcoverage} and the visibility amplitudes and phases of the self-calibrated data and the CLEAN model as functions of $(u,v)$-distance in Figure~\ref{fig:radplot}.

%% file: analysis.tex
\subsection{Kinematic Analysis of the 3C 84 Jet utilizing WISE}
\label{sec:analysis:kinematics}

We utilized a recently developed jet kinematic analysis method called WISE (Wavelet-based Image Segmentation and Evaluation; \citealt{ML2015}) to trace the position of the knot throughout the observed period of the VLBA 43 GHz monitoring data. WISE is uniquely advantageous for investigating the jet kinematics of sources characterized by complex jet structures, including M87 \citep{Mertens2016, Park2019b} and 3C 84 \citep{Hodgson2021}. Conventional kinematic methods \citep[e.g.,][]{Jorstad2017, Lister2018}, reliant on describing the jet using a compilation of components with two-dimensional Gaussian distributed brightness, encounter difficulties when applied to such sources. 

To start, we convolved the CLEAN maps using a standardized synthesized beam of $0.34\times0.16$ mas at a position angle of $0^\circ$. This beam size represents the average of the synthesized beams across the entire dataset. The alignment of the images was conducted by assuming the western component of the double or triple nuclei observed in the epochs under investigation as the core, which was presumed to be stationary over time. This assumption is supported by the following evidence: (i) the spectral index map (Figure~\ref{fig:spix}) reveals an inverted spectrum ($\alpha\sim1.3$; $S_\nu \propto \nu^\alpha$) in the assumed core position, consistent with the flat or inverted spectral index of the core in other epochs with different geometries \citep{Paraschos2021, Paraschos2022}; (ii) the core intensity remains relatively constant ($\sim2$ Jy per beam) throughout the entire observation period, while the knot intensity gradually decreases from approximately $\sim2$ to $\sim0.2$ Jy per beam (Figure~\ref{fig:intensity}); and (iii) attempting to align the images by assuming the knot's stationarity resulted in an apparent gradual motion of the downstream jet towards the west. With respect to the third line of evidence, we performed a two-dimensional cross-correlation analysis, which provides further support for our assertion that the western component represents a stationary core. The detailed results of this analysis are presented in Appendix~\ref{appendix:2dcc}.

\begin{figure}[t]
\centering
\includegraphics[width=\linewidth]{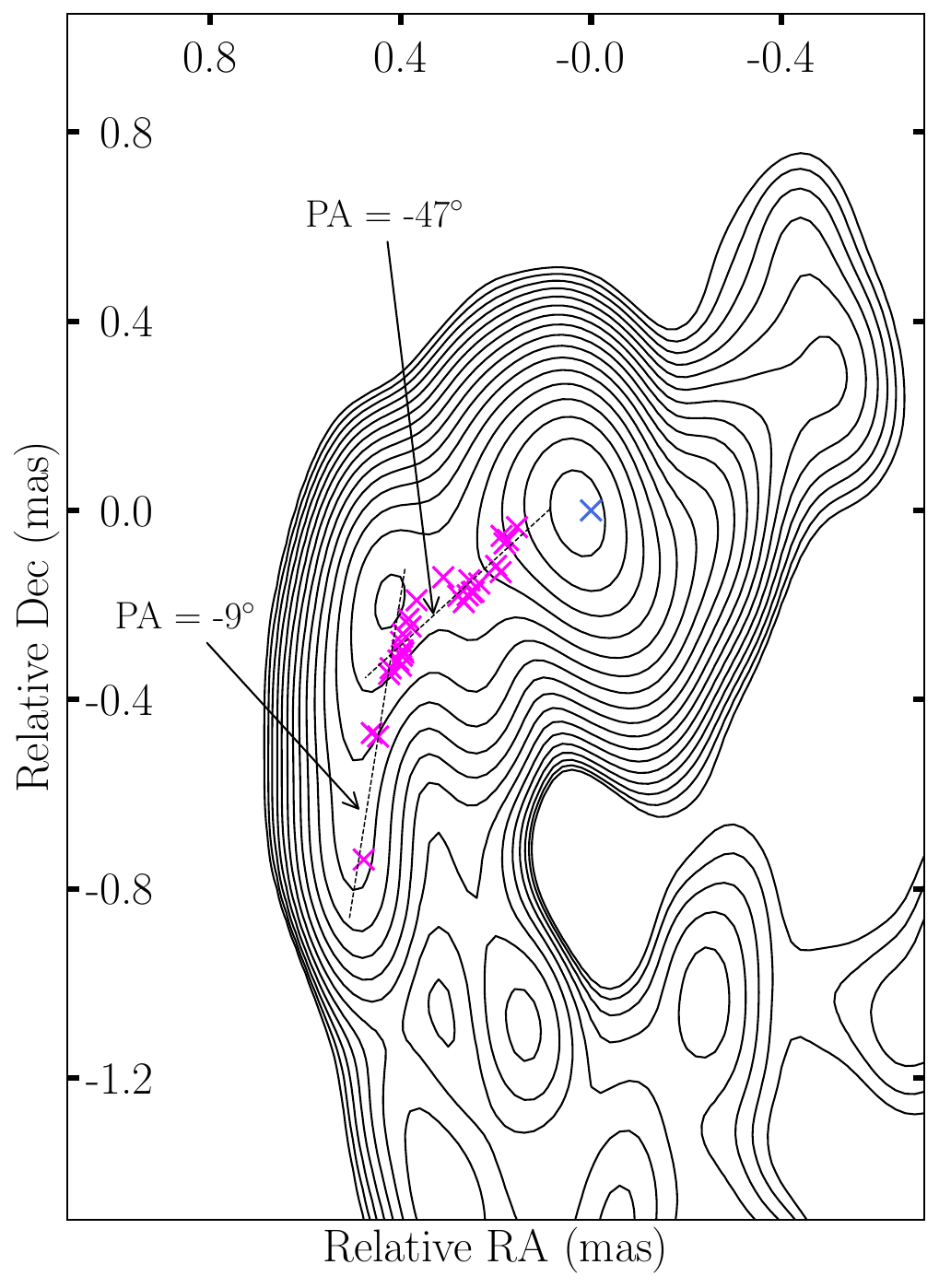}
\caption{Contours represent the total intensity image of 3C 84 observed at 43 GHz with the VLBA on November 1, 2022. The blue cross indicates the assumed position of the core, coinciding with the map origin. The magenta crosses indicate the positions of the knot, which underwent ejection from the core in the late 2010s \citep{Punsly2021} and displayed a notable variation in its apparent position angle during the early 2020s.}
\label{fig:trajectory}
\end{figure}

We employed WISE to detect significant structural patterns (SSPs) using a $3\sigma$ detection threshold on scales of 0.04, 0.06, 0.08, 0.12, 0.16, 0.24, 0.32, and 0.48 mas, utilizing both the segmented wavelet decomposition (SWD) and intermediate wavelet decomposition (IWD) methods \citep{Mertens2016}. SSPs were identified across adjacent epochs using the multiscale cross-correlation (MCC) method with a tolerance factor of 4.0 and a correlation threshold of 0.6. To accurately capture the rapid motion of the knot between the last two epochs, which had a separation of approximately 2.4 months (longer than the typical time gap between adjacent epochs, with a median separation of about 1 month), we employed a relatively large tolerance factor of 4.0 and search windows of [$-2$, $+2$] and [$-5$, $+0.03$] mas yr$^{-1}$ in the right ascension and declination directions, respectively. Notably, the absence of this setup resulted in the detected knot position showing a sudden inward motion with an apparent speed of $\beta_{\rm app} \sim -0.84$ between the last two epochs, which we attribute to the misidentification of the SSPs. Importantly, the use of the larger tolerance factor and highly asymmetric search window in the declination direction does not affect the identification of the knot in other epochs. 

\begin{figure*}[t]
\centering
\includegraphics[width=0.55\linewidth]{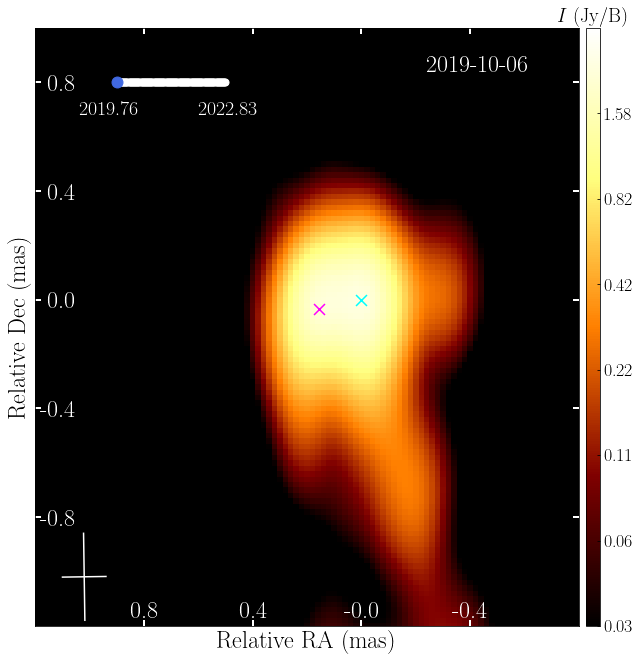}
\includegraphics[width=0.43\linewidth]{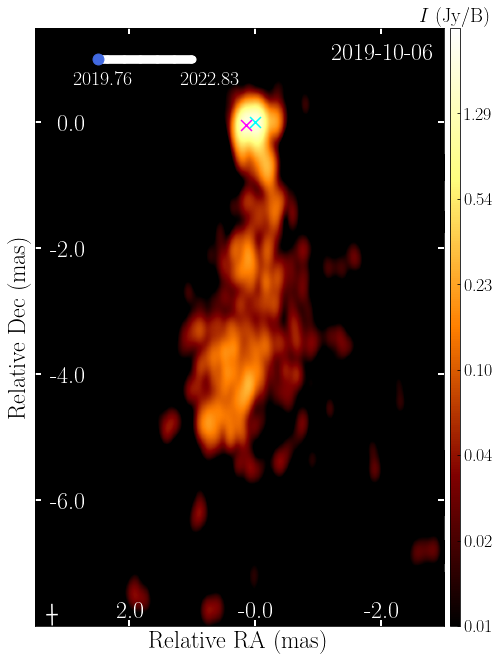}
\caption{These figures represent a single frame from the movie showcasing the zoom-in version of the core region (left; \url{https://www.dropbox.com/s/pae46qy6xsbzmvu/animated.startdelay.gif?dl=0}) and the entire jet region (right; \url{https://www.dropbox.com/s/6p0da5ugwgfv9be/wholejet.animated.startdelay.gif?dl=0}). The cyan cross indicates the assumed position of the core, whereas the magenta crosses indicate the positions of the knot identified by WISE. The top left corner displays a white bar indicating the monitoring period, while the blue circle within the bar denotes the time of observation in each epoch. The bottom left corner features white crosses representing the synthesized beam size used for convolving the individual epoch images.}
\label{fig:movie}
\end{figure*}

\begin{figure*}[t]
\centering
\includegraphics[width=\linewidth]{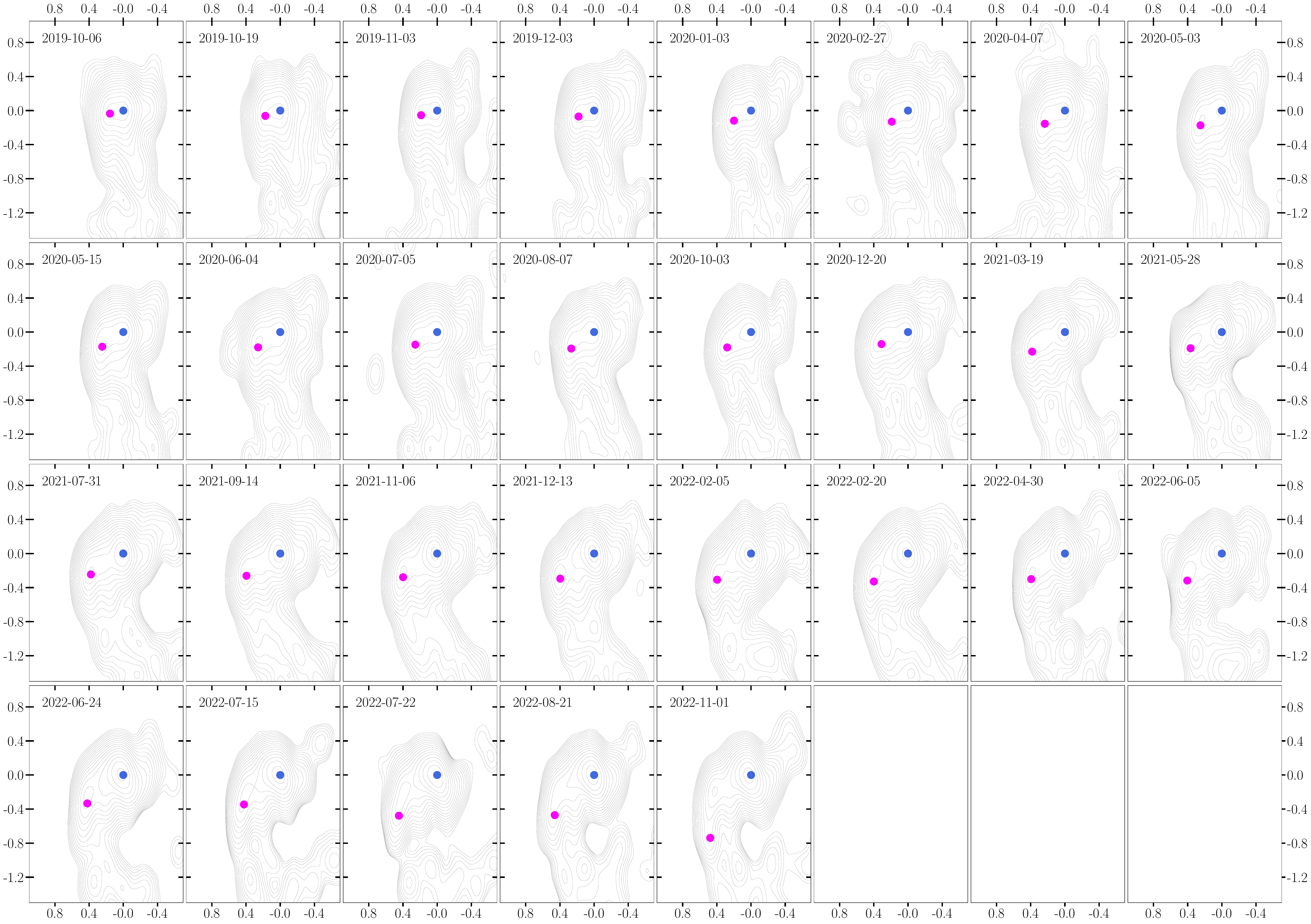}
\caption{Contour maps depicting the core region of 3C 84 based on the 43 GHz VLBA monitoring data. The positions of the assumed core, located at the map origin in each epoch, are indicated by blue filled circles. Additionally, the positions of the knots identified through the WISE analysis are marked by magenta filled circles. The observation date for each epoch is indicated in the top left corner of the respective image.}
\label{fig:composite}
\end{figure*}

Our analysis revealed successful tracing of the knot at a scale of 0.04 mas. However, at larger scales, it appears that the knot is not adequately separated from neighboring jet structures. Figure~\ref{fig:trajectory} displays the trajectory of the knot as determined by WISE. Initially, the knot exhibited eastward motion. However, between late 2021 and early 2022, the knot underwent a significant deflection, resulting in a change of its propagation direction. The position angle shifted from approximately $-47^\circ$ to $-9^\circ$. Following this deflection\footnote{It is noteworthy that the trajectory of the knot may display a helical appearance in earlier epochs; however, this pattern does not appear to endure in the later epochs. The trajectory of a helical jet, when projected onto the sky, manifests as a quasi-sinusoidal transverse oscillation relative to the central axis, as demonstrated in previous studies \citep[e.g.,][]{Kun2014}. This characteristic stands in contrast to the knot's continuous southerly motion following its deflection. Consequently, we assert that the data favor the deflection scenario over the interpretation of helical motion.}, the direction became consistent with that of the parsec-scale jet. We created movies showcasing the 3C 84 jet using the VLBA 43 GHz observations, including the identified knot position. The first movie provides a zoom-in version of the core region and can be accessed at the following link: \url{https://www.dropbox.com/s/pae46qy6xsbzmvu/animated.startdelay.gif?dl=0}. The second movie covers the entire jet region and can be viewed here: \url{https://www.dropbox.com/s/6p0da5ugwgfv9be/wholejet.animated.startdelay.gif?dl=0}. Based on the movies, it is evident that WISE successfully tracked the position of the knot. Figure~\ref{fig:movie} displays a single frame from each movie. Figure~\ref{fig:composite} depicts contour maps of the core region of 3C 84, with the positions of the knots identified by WISE superimposed.

\subsection{High-resolution Spectral Index Mapping with GVA}
\label{sec:analysis:spix}

%The GVA offers extensive baselines and achieves a densely filled $(u,v)$-coverage (Figure~\ref{fig:uvcoverage}). When employing uniform weighting of the visibility data, the resulting synthesized beam shape is $\sim0.36\times0.17$ mas, oriented at a position angle of roughly $18^\circ$. This is comparable to the synthesized beam from the last epoch data of the VLBA 43 GHz observations, which yielded a beam shape of $\sim0.31\times0.15$ mas at a position angle of around $6^\circ$. It is worth noting that the VLBA observation took place approximately one week prior to the GVA 22 GHz observation.

The inclusion of multiple telescopes in the GVA 22 GHz observation provided excellent $(u,v)$-coverage (Figure~\ref{fig:uvcoverage}) and allowed for very high angular resolution, resulting in a synthesized beam of approximately $0.35\times0.16$ milliarcseconds (mas) at a position angle of approximately $18^\circ$. The utilization of very long baselines, with lengths of up to nearly $1\ \rm G\lambda$, contributed to the enhanced resolution. To generate a spectral index map, we utilized the image from this data in conjunction with the VLBA 43 GHz image obtained approximately one week earlier (November 1, 2022). Since the shapes of the synthesized beams from the two images closely resembled each other, we achieved a high-quality spectral index map without significant artifacts arising from different angular resolutions (\citealt{Hovatta2014}). Color maps of the GVA 22 GHz image and the VLBA 43 GHz image are displayed in Appendix~\ref{appendix:gva}.

\begin{figure*}[t]
\centering
\includegraphics[width=0.8\linewidth]{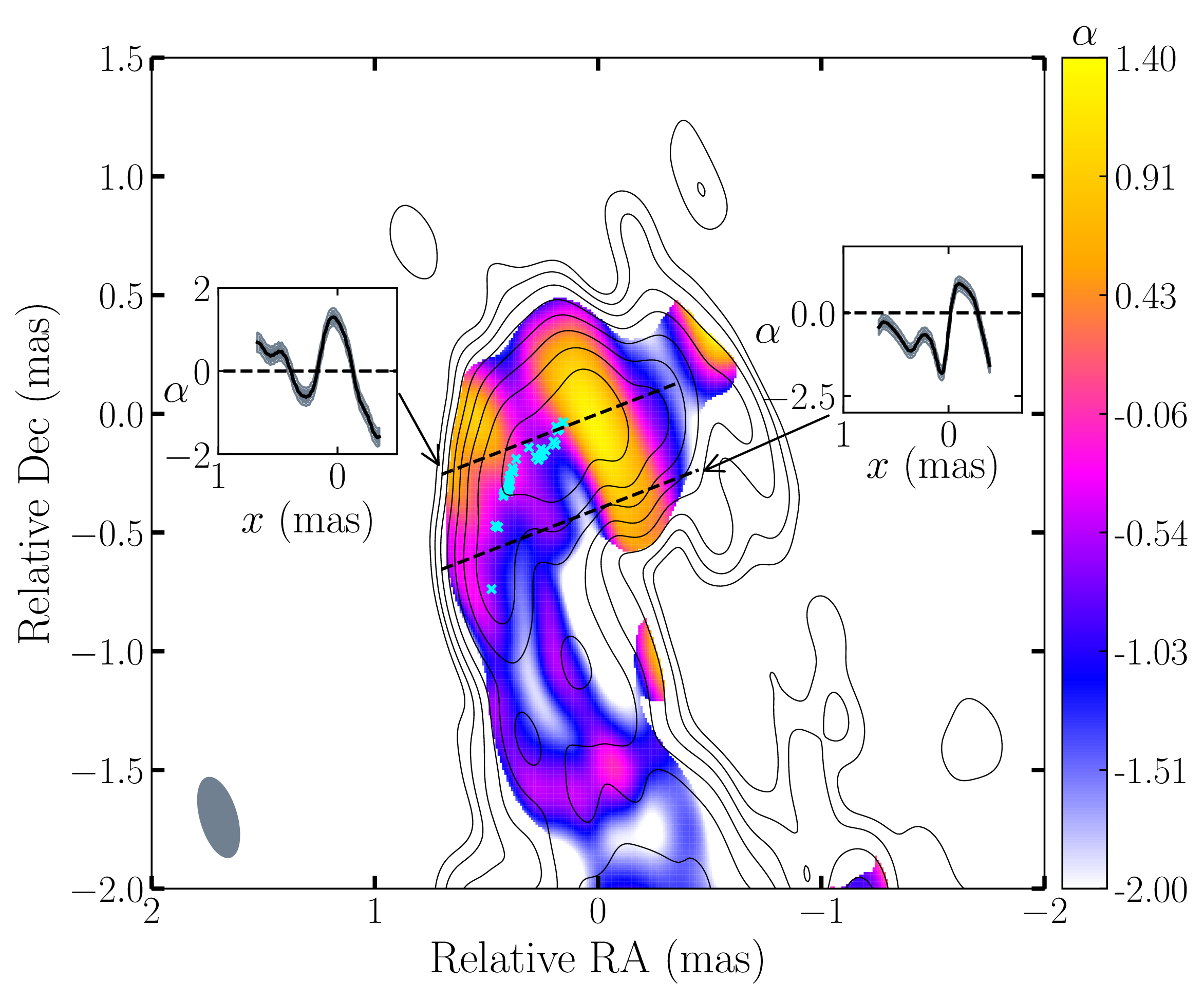}
\caption{Color represents the spectral index distribution of 3C 84 derived from the images obtained with the GVA at 22 GHz on November 9, 2022, and the VLBA at 43 GHz on November 1, 2022, overlaid on the total intensity contours at 22 GHz. Contours initiate at a level of five times the off-source image rms noise and increment by a factor of two. Prior to computing the spectral index, the images were aligned to maximize the cross-correlation coefficient of the optically thin jet emission. The inset figures depict the spectral indices along the slices indicated by the black dashed lines. The cyan crosses indicate the positions of the knot identified through the WISE analysis. The jet edge near the deflection point of the knot exhibits an inverted spectrum, with a maximum spectral index of $\alpha_{\rm inv}\sim0.95$. This inverted spectrum is likely attributed to the presence of a dense cold ambient medium, which is believed to contribute to the observed deflection of the knot. A corresponding plot illustrating the entire parsec-scale jet region can be seen in Figure~\ref{fig:spix_entire}.}
\label{fig:spix}
\end{figure*}

To generate a spectral index map, we utilized the two frequency images obtained by convolving the CLEAN models in the images with the larger synthesized beam from the GVA data. Comparing VLBI images obtained at different frequency bands can be challenging due to the lack of absolute phase information and the core-shift effect typically observed in AGN jets \citep{Park2021b}. To address these issues, we performed a two-dimensional cross-correlation on the optically thin jet emission within the images at a declination of $<-1.5$ mas \citep{CG2008}. The region employed for cross-correlation computation is delineated by a black dashed box in Figure~\ref{fig:spix_entire}. Through this analysis, we identified the optimal alignment by shifting the low-frequency image by $-0.01$ mas in right ascension and $+0.01$ mas in declination\footnote{It is worth noting that the core-shift between 22 and 43 GHz, as determined in this study, appears to be smaller compared to the trend observed in a previous investigation that utilized images at 15, 43, and 86 GHz \citep{Paraschos2021}. However, considering the complex evolution of the source and the active ejection of new knots, it is possible for the core-shift to exhibit time variability, as observed in other AGN jets \citep{Lisakov2017}. Furthermore, the core-shift derived between 5 and 15 GHz in 3C 84 \citep{Savolainen2023} is notably smaller than what was expected based on the previous study \citep{Paraschos2021}.}. The spectral index map was then derived based on this alignment. Although the synthesized beam shapes at both frequencies are similar, the difference in $(u,v)$-coverage can still impact the spectral index map, particularly in regions with low signal-to-noise ratios (SNRs; \citealt{Hovatta2014}). To mitigate this effect and obtain a conservative spectral index map, we derived the spectral index only for pixels where the observed total intensity exceeds 20 times the off-source image rms-noise levels at both frequencies. Spectral index errors were computed following the approach outlined in \cite{KT2014}. To account for uncertainties in the absolute amplitudes of the visibility data, we introduced additional systematic errors by including 10\% amplitude errors for each pixel in every image, with calculations performed in quadrature. The distribution of spectral index errors is presented in Appendix~\ref{appendix:spixerr}.

The resulting spectral index map is presented in Figure~\ref{fig:spix}. The core, serving as a reference point for image alignment across different epochs, displays a highly inverted spectrum with a spectral index $\alpha\sim1.3$, defined as $I_\nu \propto \nu^\alpha$, where $I_\nu$ represents the intensity at frequency $\nu$. This finding aligns well with previous VLBI observations that also reported inverted or flat spectra in the core \citep{Paraschos2021, Paraschos2022}. Interestingly, we observed a similar behavior in the jet edge region near the location where the knot deflection was observed. Hereafter, we will refer to it as 'the knot deflection point.' This region exhibits inverted spectra with $\alpha_{\rm inv}\sim0.95$ near the deflection point, contrasting with the downstream portion of the jet, which mostly displays steep spectra with $\alpha < 0$. In Appendix~\ref{appendix:beam}, we demonstrate that the inverted spectral region persists even with larger restoring beam sizes, albeit with reduced contrast. We also conducted a test to investigate the persistence of the inverted spectral region when deliberately introducing artificial image misalignment between the 22 and 43 GHz images. The primary objective of this analysis is to affirm the robustness of the conclusions presented in our paper concerning potential image alignment artifacts arising from imperfect alignment procedures. The results are presented in Appendix~\ref{appendix:misalignment}, and we found that the inverted spectral region remains observable even in the presence of substantial image misalignment. Additionally, we conducted imaging simulations to investigate the possibility that the inverted spectral region might arise as an artifact due to factors such as differences in $(u,v)$-coverages between the 22 and 43 GHz data, as well as imperfections in antenna gain calibration and imaging procedures. The details of our simulation methodology and the results are outlined in Appendix~\ref{appendix:synthetic}. Based on this analysis, we conclude that the observed inverted spectral region is not an artifact but rather an authentic signature inherent to the source.

\begin{figure}[t]
\centering
\includegraphics[width=\linewidth]{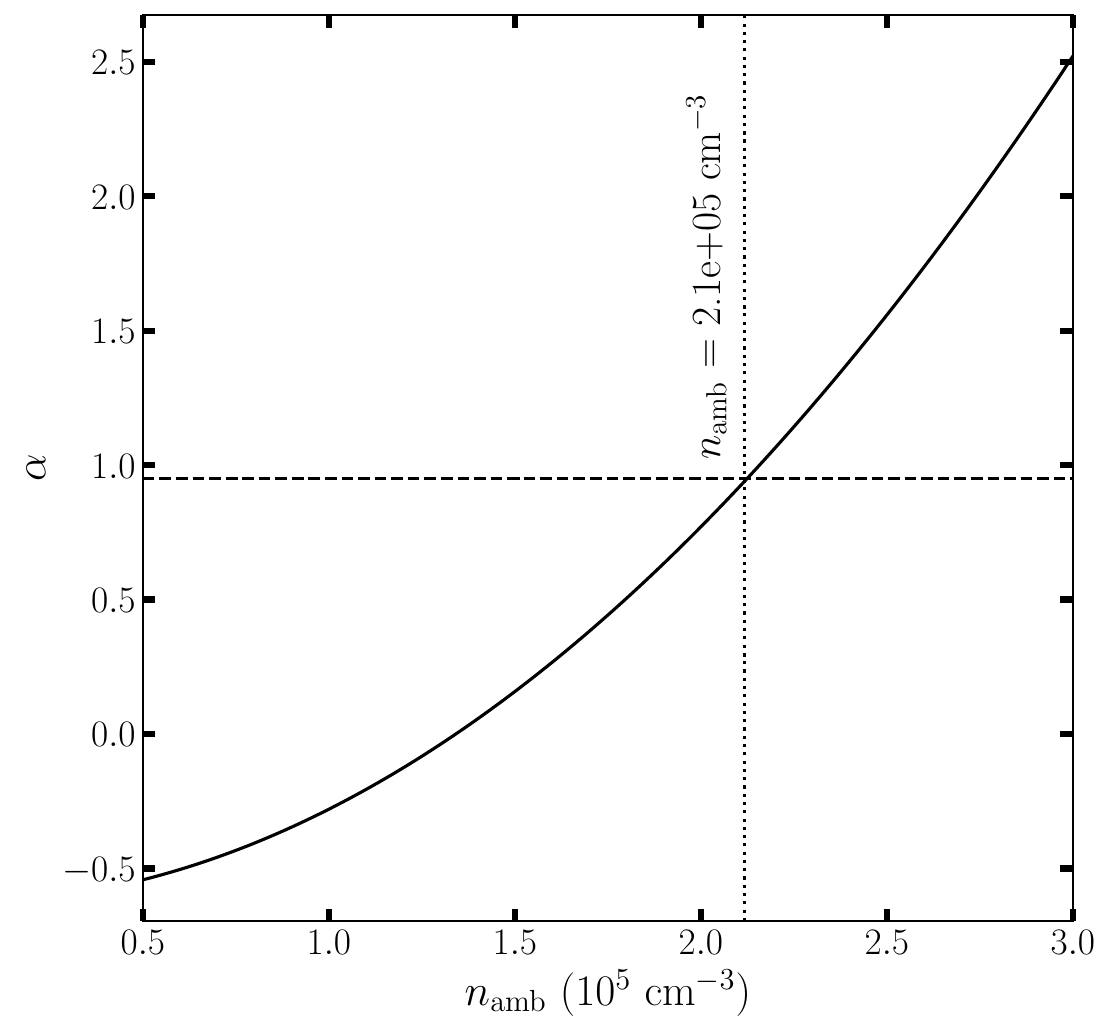}
\caption{Spectral index $\alpha$ between 22 and 43 GHz expected to be produced by FFA due to the ambient medium with density $n_{\rm amb}$. The density of $n_{\rm amb} = 2.1\times10^5{\rm\ cm^{-3}}$ is required to reproduce the observed inverted spectrum with $\alpha_{\rm inv} \sim 0.95$ near the deflection point.}
\label{fig:alpha}
\end{figure}

The 3C 84 jet is known for its active interaction with an ambient medium on parsec scales. When there are indications of jet-medium collisions, the interaction with the ambient medium frequently results in flat or inverted spectra within the jet, primarily caused by free-free absorption (FFA; \citealt{Kino2021}). By assuming that the observed flat spectrum with $\alpha_{\rm inv}\sim0.95$ near the deflection region is attributed to FFA by the ambient medium\footnote{We acknowledge that the observed inverted spectrum could be affected by the fact that different parts of the jet might be visible at varying frequencies. For example, Perlman et al. (1999), through their study of the polarization differences in the M87 jet using optical and radio wavelengths, suggested that the jet layer closer to the axis contributes more to higher frequency emissions, while the layer near the jet boundary contributes more to lower frequency emissions. However, if the 43 GHz image were to display a narrower section than the 22 GHz image, it would lead to a steep spectrum at the jet's edges, contradicting our actual observations. Furthermore, in this scenario, the question remains: "Why did we only observe the inverted spectrum at the edge near the point where the knot deflects?"} (refer to Section~\ref{sec:discussion:ssa} for a discussion on the possibility of synchrotron-self absorption), we can deduce the density of the medium. The free-free optical depth can be expressed as $\tau_{\rm ff} \approx 5.6\times10^{-8}\bar{g}(T/10^4{\rm K})^{-3/2}(n_{\rm amb}/{\rm cm^{-3}})^2(\nu / {\rm GHz})^{-2}(L / {\rm pc})$, assuming a uniform density pure hydrogen plasma \citep{Levinson1995}. Here, $T$, $n_{\rm amb}$, and $L$ represent the temperature, electron density, and size of the absorbing medium, respectively. We assume a temperature of $T=10^4\ K$ \citep{Levinson1995, FN2017, Wajima2020} and estimate $L \sim 0.070\ {\rm pc}$ based on the size of the region exhibiting inverted spectra. The thermal average Gaunt factor, $\bar{g}$, is approximately 4 for $T\sim10^4{\rm \ K}$ and $\nu\sim22$--43 GHz \citep{FN2017}. We further assume that the observed inverted spectrum is generated by the FFA of the background optically thin synchrotron emission, characterized by a spectral index of $\alpha_{\rm steep}\sim-0.63$. This spectral index is obtained from the optically thin spectral region between the jet edge near the knot deflection point and the core (Figure~\ref{fig:spix}). With this information, we can compute the expected spectral index between 22 and 43 GHz for a given electron density of the ambient medium. Figure~\ref{fig:alpha} shows the derived spectral index as a function of $n_{\rm amb}$. To reproduce the observed inverted spectrum with $\alpha_{\rm inv} \sim 0.95$, an electron density of $n_{\rm amb} \sim 2.1\times10^5\ {\rm cm^{-3}}$ is required for the ambient medium.

\subsection{Speed of the Knot}
\label{sec:analysis:speed}

\begin{figure}[t]
\centering
\includegraphics[width=\linewidth]{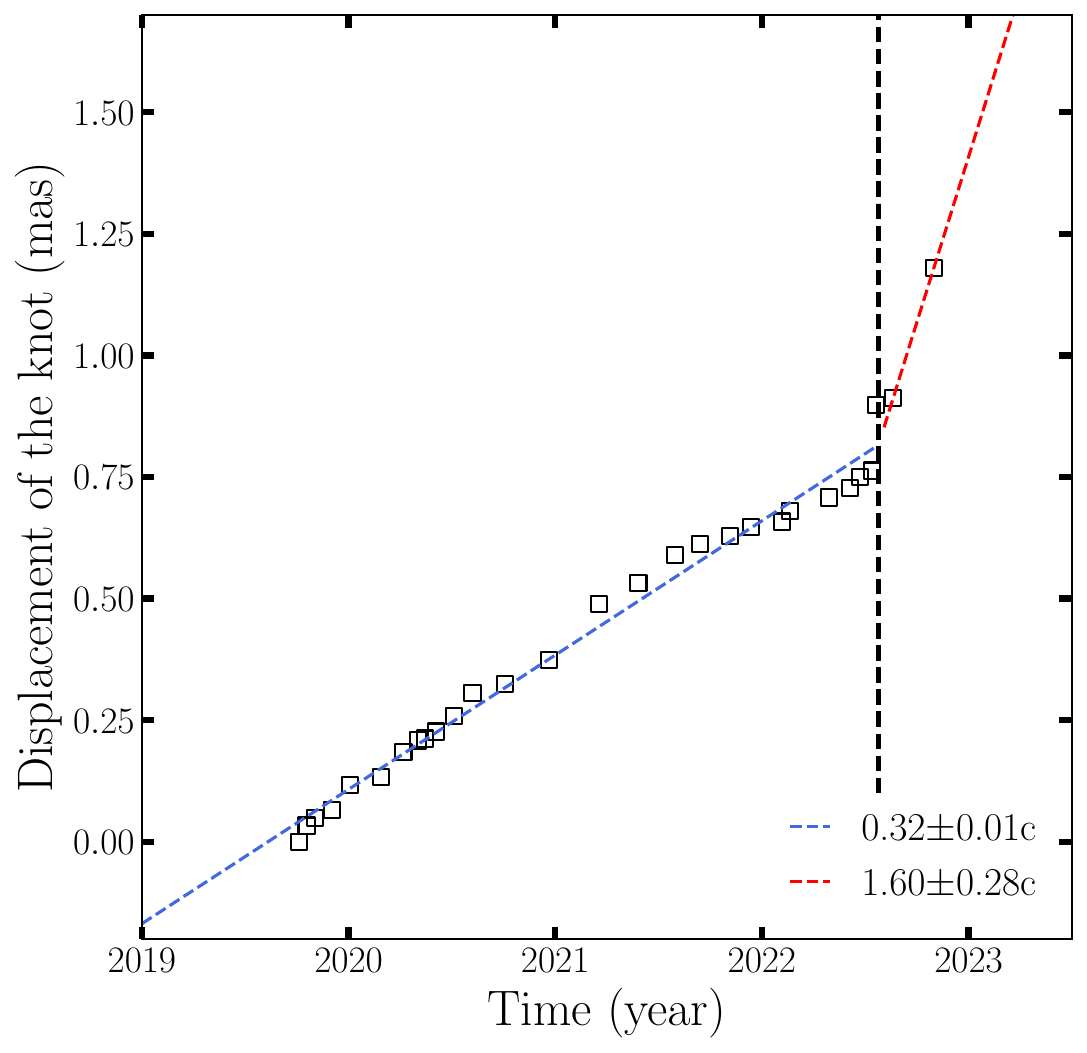}
\caption{The relative displacement of the knot position over time is depicted. Initially, the knot gradually moves outward with an apparent speed of $\beta_{\rm app} = 0.32\pm0.01$ until mid-2022. However, subsequent to mid-2022, the knot exhibits a significantly accelerated motion, with an apparent speed of $\beta_{\rm app} = 1.60\pm0.28$.}
\label{fig:separation}
\end{figure}

The trajectory of the knot does not follow a simple ballistic path; instead, it exhibits a complex behavior (Figure~\ref{fig:trajectory}). Therefore, determining the apparent speed of the knot solely based on its separation from the core is not feasible. To analyze the knot's motion, we employed a third-order polynomial function to fit its two-dimensional trajectory. We note that a second-order polynomial function did not fit the data well, and the results of applying higher order polynomial functions were not significantly different from those of the third-order polynomial function. The relative displacement of the knot position along this function was computed as a function of time, as depicted in Figure~\ref{fig:separation}. Interestingly, we observed a distinct change in the slope of the displacement around mid-2022. To capture this behavior, we employed a piecewise linear function comprising two linear segments with different slopes that connect at a specific time. This approach allowed us to identify a sudden increase in the apparent speed of the knot from $\beta_{\rm app} = 0.32\pm0.01$ to $1.60\pm0.28$ during mid-2022. It is not straightforward to pinpoint the exact time of the knot deflection, so it is unclear if the observed abrupt acceleration is coincident with the deflection.

%% file: discussion.tex
\subsection{Exploring the Origin of the Inverted Spectrum near the Knot Deflection Point: FFA vs. SSA}
\label{sec:discussion:ssa}

In our analysis of the observed inverted spectrum near the knot deflection point, we initially attribute the phenomenon to FFA resulting from the dense ambient medium (Section~\ref{sec:analysis:spix}). However, an alternative explanation could be Synchrotron Self-Absorption (SSA) caused by the compression of the jet by the ambient medium \citep{Mimica2009, Hovatta2014}, which is also consistent with our main claim that the abrupt knot deflection observed in this study is due to the dense ambient medium (see Section~\ref{sec:discussion:ambient}). Nevertheless, in this section, we present a detailed discussion on why the FFA scenario is more favored over the SSA scenario, based on the energy density ratio between synchrotron emitting particles ($u_{\rm p}$) and the magnetic field ($u_{\rm B}$) in the jet frame. 

Building upon previous studies \citep{Kim2018}, the energy density ratio can be derived as follows: $u_{\rm p} / u_{\rm B} = (T_{\rm B, eq}/T_B)^{-17/2}$ \citep{Readhead1994}. Here, $T_B$ represents the intrinsic brightness temperature in the source rest frame for an emitting component with a circular Gaussian source brightness distribution, given by: $T_B = 1.22\times10^{12}S(1+z)/\nu^2d^2\delta \ {\rm K}$. In this equation, $S$ denotes the component flux density in Jy at frequency $\nu$ in GHz, $d$ represents the full width at half-maximum of the component in units of mas, $z=0.0176$ is the redshift of the source, and $\delta = [\Gamma(1 - \beta\cos\theta_{\rm view})]^{-1}$ denotes the Doppler factor. Here, $\Gamma = (1-\beta^2)^{-1/2}$ represents the Lorentz factor, $\beta$ denotes the intrinsic jet speed in units of the speed of light, and $\theta_{\rm view}$ represents the viewing angle. The equipartition brightness temperature is given by: $T_{\rm B, eq} = t(\alpha)10^{11}[(s/{\rm pc})(\nu_{\rm turn}/{\rm GHz})^{1.5+\alpha}]^{1/8}\ {\rm K}$. This temperature is expected when the energy density ratio is unity. Here, $\alpha$ represents the spectral index of the optically thin part of the spectrum, $t(\alpha)$ is a numerical factor dependent on $\alpha$, $s$ denotes the linear size of the emitting region in parsec, and $\nu_{\rm turn}$ is the synchrotron turnover frequency in GHz.

\begin{figure}[t]
\centering
\includegraphics[width=\linewidth]{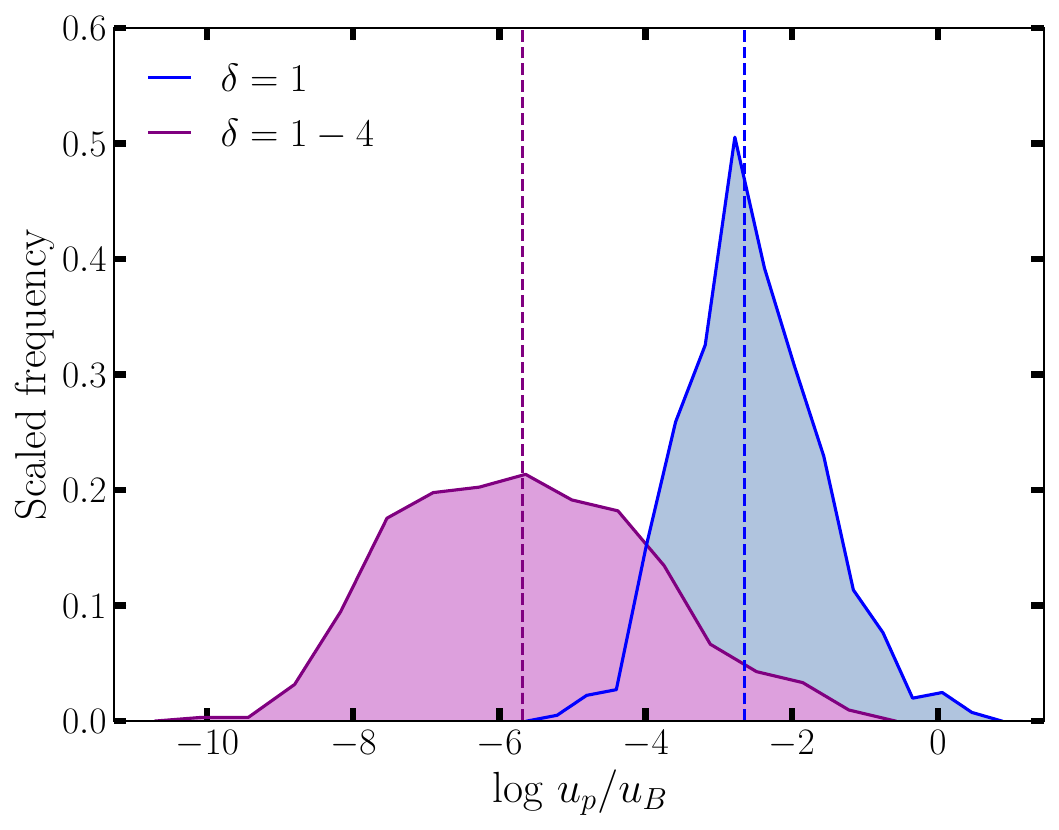}
\caption{Distribution of $\log u_{\rm p} / u_{\rm B}$ obtained from the Monte Carlo sampling of the individual quantities comprising the ratio. The blue and magenta areas show the distributions for the case of $\delta=1$ and $\delta=$1--4, respectively. The vertical dashed lines indicate the position of the distribution medium at approximately $\sim-5.7$ (magenta) and $\sim-2.6$ (blue).}
\label{fig:ratio}
\end{figure}

\begin{figure*}[t]
\centering
\includegraphics[width=0.8\linewidth]{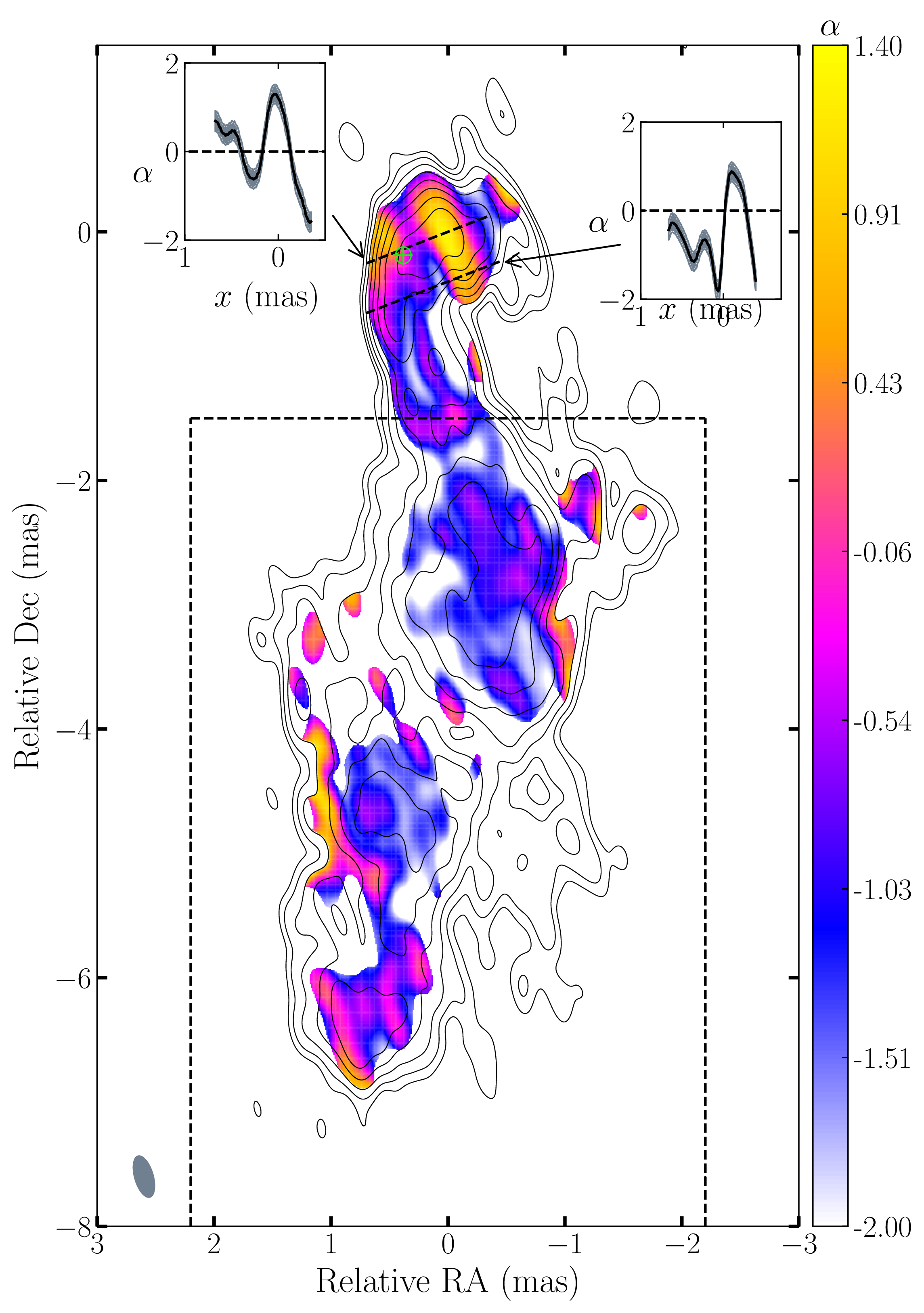}
\caption{Similar to Figure~\ref{fig:spix}, but displaying the complete parsec-scale jet region. The \texttt{modelfit} component, located near the inverted spectrum region close to the jet deflection point and used to discuss the possibility of SSA for the observed inverted spectrum, is represented by the encircled lime-colored cross in the figure. The optically thin jet region utilized for cross-correlation calculations in Section~\ref{sec:analysis:spix} is demarcated by a black dashed box.}
\label{fig:spix_entire}
\end{figure*}

\begin{figure*}[t]
\centering
\includegraphics[width=0.47\linewidth]{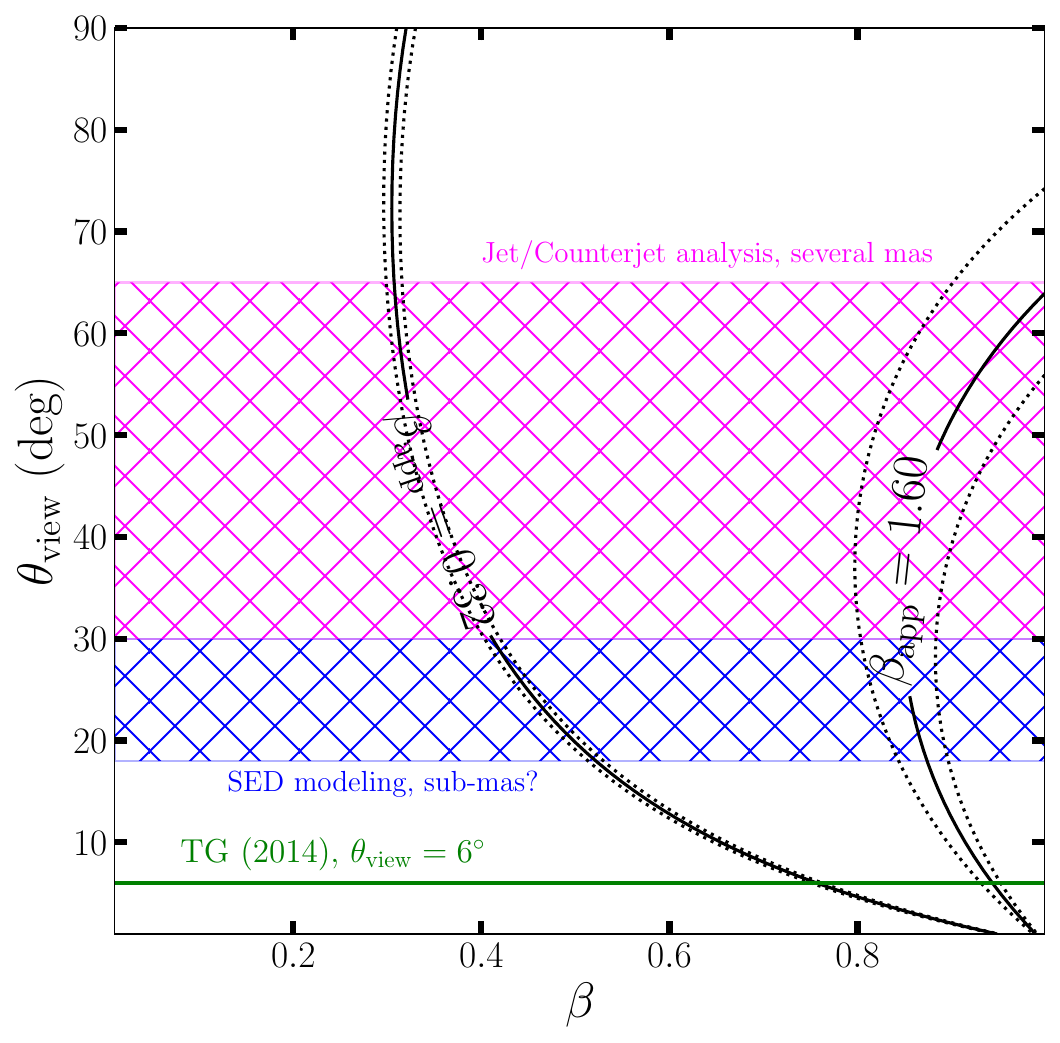}
\includegraphics[width=0.482\linewidth]{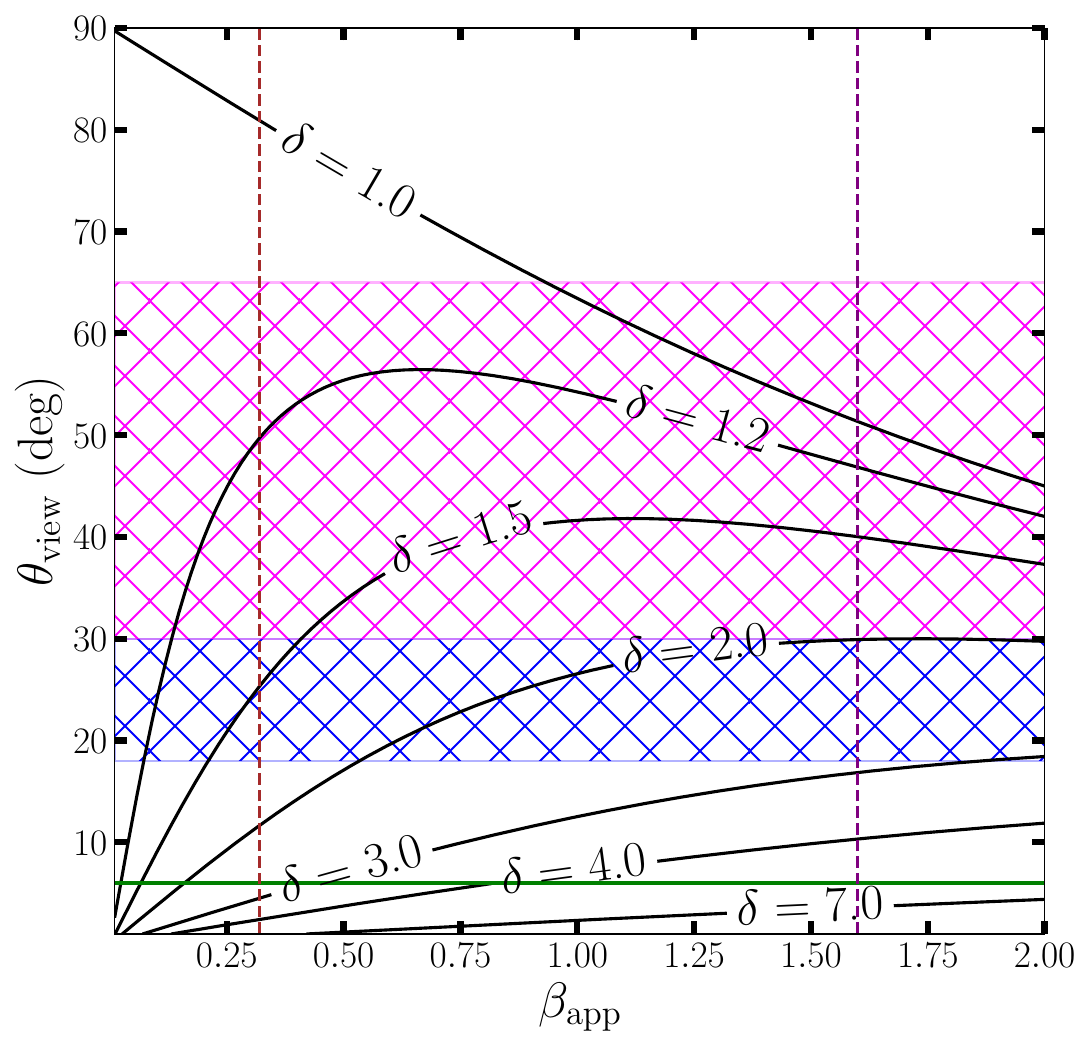}
\caption{\textit{Left:} the figure presents contours denoting $\beta_{\rm app} = 0.32$ and $1.60$ as a function of $\beta$ and $\theta_{\rm view}$, represented by solid black lines. The contours corresponding to the apparent speed values deviated by $1\sigma$ errors are depicted by dotted black lines. The blue hashed region indicates the inferred range of jet viewing angles, $\theta_{\rm view} = 18-30^\circ$, based on previous studies employing SED modeling \citep{Abdo2009, Aleksic2014, TG2014}. Similarly, the red hashed region represents the range of $\theta_{\rm view} = 30-65^\circ$ inferred from observations of the counterjet \citep{Walker1994, Asada2006, FN2017}. The green solid line emphasizes a viewing angle of $6^\circ$, which yielded a satisfactory fit to the observed SED but was considered unreliable due to its significant deviation from the viewing angles determined through radio observations \citep{TG2014}. \textit{Right:} Contours corresponding to various values of the Doppler factor $\delta$ are presented as functions of $\beta_{\rm app}$ and $\theta_{\rm view}$. The vertical dashed lines indicate the observed apparent speeds of the knot, $\beta_{\rm app} = 0.32$ and $1.60$.}
\label{fig:speed}
\end{figure*}

\begin{figure}[t]
\centering
\includegraphics[width=\linewidth]{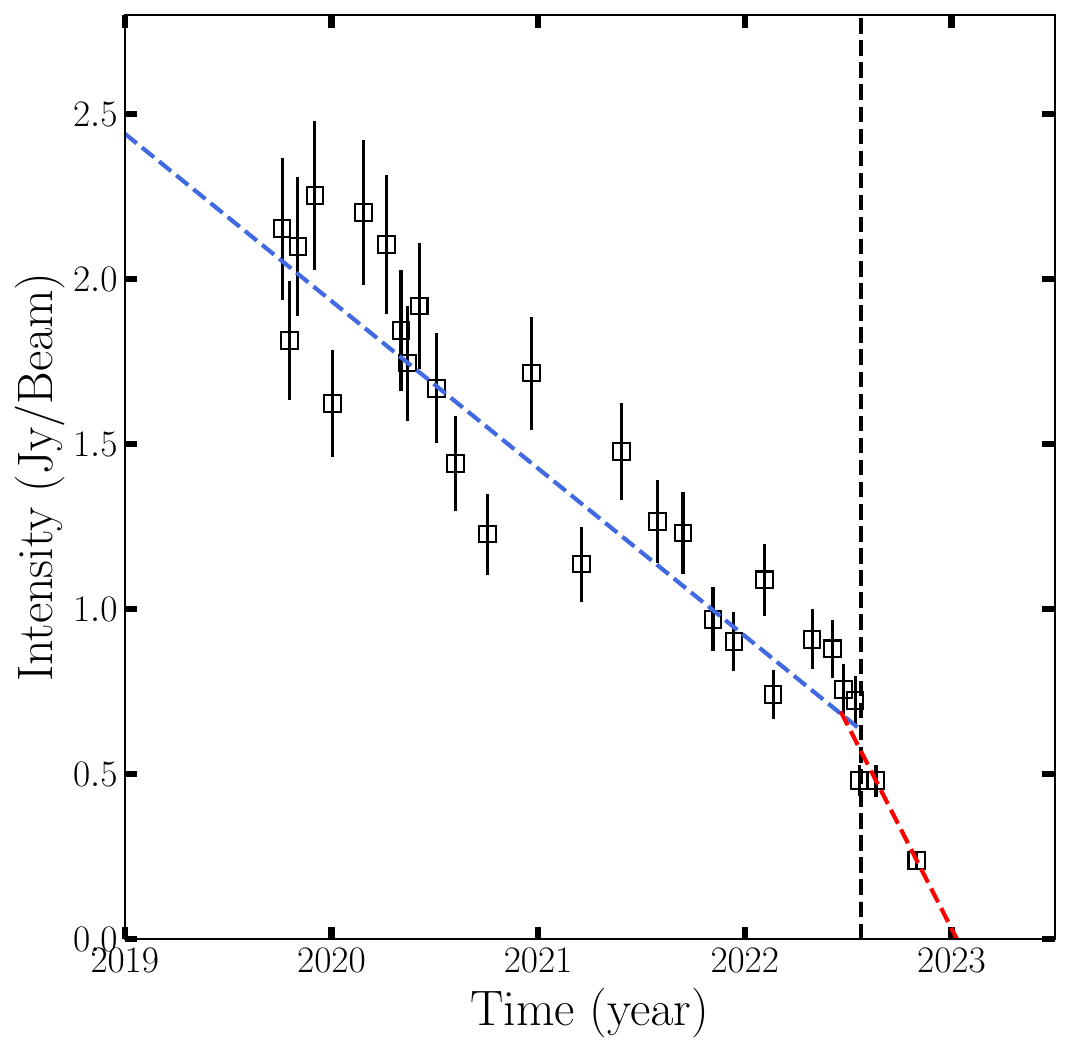}
\caption{The total intensity of the knot, measured in Jy per beam, is plotted against time. The uncertainties are obtained by combining the image rms errors and the 10\% systematic errors arising from the uncertainties in visibility amplitudes through quadrature. The blue and red dashed lines represent the best-fit linear functions for the observed intensities prior to and after mid-2022 (indicated by the black vertical dashed line), respectively. This transition coincides with a sudden acceleration in the knot's apparent speed.}
\label{fig:intensity}
\end{figure}
 
To determine the intrinsic brightness temperature and the equipartition temperature near the deflection point, we utilized the visibility data obtained from the VLBA 43 GHz observation conducted on November 1, 2022. The data was analyzed using the \texttt{modelfit} task in Difmap, employing a set of circular Gaussian components to fit the observed data. Our analysis specifically focused on the \texttt{modelfit} component that was closest to the region displaying the inverted spectrum near the knot deflection point (refer to Figure~\ref{fig:spix_entire} for detailed information regarding this component). From this component, we obtained the values of $S = 0.98\ \rm Jy$ and $d = 0.14\ \rm mas$, which are used to derive $T_B$. 

To determine the turnover frequency $\nu_{\rm turn}$, we employed the specific intensity formula expected for synchrotron radiation \citep{Fromm2013}. This formula is given by $I_\nu = I_{\nu_1}(\nu/\nu_1)^{\alpha_t}[1-\exp(-(\nu/\nu_1)^{\alpha_{\rm steep} - \alpha_t})]$, where $\nu_1$ is the frequency where the synchrotron optical depth becomes unity, $\alpha_t$ is assumed to be 5/2, which is expected for a homogeneous source. We set the optically thin spectral index $\alpha_{\rm steep}$ to $-0.63$ based on the observed spectral index between the inverted spectrum region and the core (as shown in Figure~\ref{fig:spix}). Additionally, the factor $t(\alpha)$ was assigned a value of $0.65$ \citep{Singal2009}. By selecting $\nu_{\rm turn}$ as $32.5\ \rm GHz$, we were able to successfully reproduce the observed spectral index of $\alpha_{\rm inv} = 0.95$ for the inverted spectrum region spanning from 22 to 43 GHz.

To address the uncertainties in various quantities involved in the estimation of the energy density ratio, we employed a Monte Carlo approach. The uncertainties in the \texttt{modelfit} component quantities were determined using an empirical relationship established in previous studies \citep{Jorstad2017}, which correlates the uncertainties with the observed brightness temperature. The uncertainty in $\nu_{\rm turn}$ was derived by considering the associated uncertainty in the spectral index. For each quantity, we performed a Monte Carlo simulation by randomly drawing 1,000 numbers from Gaussian distributions. The standard deviations of these distributions were set to match the corresponding uncertainties. By propagating these uncertainties through the calculations, we obtained the expected distribution of $\log(u_{\rm p} / u_{\rm B})$. Estimating the uncertainties in $\delta$ was challenging due to the uncertainty in the intrinsic jet speed and viewing angle (see Section~\ref{sec:discussion:acceleration}). We considered two scenarios: one with $\delta = 1$, assuming no uncertainty, and another with a uniform distribution of $\delta$ ranging from 1 to 4.

We present the probability distribution of $\log(u_{\rm p} / u_{\rm B})$ in Figure~\ref{fig:ratio}. The distribution exhibits peaks at $u_{\rm p} / u_{\rm B}\sim2\times10^{-3}$ and $\sim2\times10^{-6}$ for $\delta=1$ and the range $\delta=1$--$4$, respectively. If the inverted spectrum near the knot deflection point is caused by SSA, it would imply that $u_{\rm B}$ dominates $u_{\rm p}$ by several orders of magnitude. When comparing this result with past studies of the energetics of other AGN jets, we find the SSA scenario less likely for 3C 84. For instance, in the well-studied case of M87, $u_{\rm B}$ dominance is suggested near the black hole \citep{Kino2015, Kim2018}, but $u_{\rm p} \gtrsim u_{\rm B}$ is suggested downstream of the M87 jet, deviating from a state of complete $u_{\rm B}$ dominance \citep{Kino2014, MAGIC2020, EHTMWL2021}. Similarly, previous studies have reported $u_{\rm p} > u_{\rm B}$ for the jets in several blazars \citep{Kino2002, Hayashida2015}. These results align with a model in which the jet accelerates through the conversion of $u_{\rm B}$ into the jet's kinetic energy, as demonstrated in GRMHD simulations \citep{McKinney2006, Chatterjee2019}. Therefore, we conclude that the observed inverted spectrum near the knot deflection point, which is at a distance scale larger than several hundred gravitational radii ($R_g$), is less likely caused by SSA. Instead, we consider FFA as a more plausible explanation for the inverted spectrum, especially considering that FFA signatures have been observed in other parts of the parsec-scale jet in 3C 84 \citep{Walker1994, Vermeulen1994, Walker2000, FN2017, Wajima2020, Kino2021}.

\subsection{The Nature of the Ambient Medium and Its Role in Shaping the Jet of 3C 84}
\label{sec:discussion:ambient}

In Section~\ref{sec:analysis:spix}, we obtained the density of the FFA medium responsible for the inverted spectrum observed near the knot deflection point, which is estimated to be $n_{\rm amb} \sim 2.1\times10^5\ {\rm cm^{-3}}$ (Figure~\ref{fig:alpha}). Notably, this density aligns well with the density of the medium believed to be responsible for the FFA of the counterjet emission \citep{Plambeck2014, FN2017, Wajima2020}. However, the inverted spectrum region is located at a separation of approximately $0.20\rm\ pc$ in the right ascension direction and within the range of $0.06$ to $0.17\rm\ pc$ in the declination direction, for jet viewing angles ranging from $18^\circ$ to $65^\circ$. These separations correspond to a ratio of height ($H$) to radius ($R$) of the putative accretion disk of $H/R \sim 0.30$ to $0.87$. In our estimation, we assume that the separation between the black hole and the core is negligible, as evidenced by the small core-shift observed between 22 and 43 GHz. The range of this ratio is too large to be explained by a standard geometrically thin accretion disk  \citep{Netzer2013}, indicating that the FFA absorbing medium may not be part of the thin disk. However, it has been proposed that the FFA in the counterjet emission arises not only from the thin disk but also from a collection of dense clumpy clouds located above the disk  \citep{Wajima2020}. The density derived for the FFA medium closely matches that of the clumpy cloud  \citep{Wajima2020}, suggesting that the cloud could be responsible for the observed FFA and the deflection of the knot. The exact nature of these clumpy clouds remains uncertain. One possibility is that they are associated with chaotic cold accretion  \citep{Gaspari2013}, a hypothesis supported by recent observations using the Atacama Large Millimeter/submillimeter Array (ALMA) on scales of less than $100$ pc in this galaxy \citep{Nagai2019}. Alternatively, these clouds could be linked to winds or outflows originating from the central active nucleus \citep{Wada2012}.

AGN jets are thought to be collimated by the pressure exerted by an external confining medium in the so-called jet acceleration and collimation zone \citep{Eichler1993, BL1994, Komissarov2007, Komissarov2009, Lyubarsky2009, Vlahakis2015, Nakamura2018}, presumably located within the jet distance of less than $\sim10^4-10^6$ $R_g$ \citep{Marscher2008}. However, the exact nature of this medium remains uncertain. In the case of nearby low-luminosity AGN such as M87, which are believed to be powered by hot accretion flows \citep{YN2014, Blandford2019}, non-relativistic gas outflows known as winds, originating from the hot accretion flows \citep{Yuan2015, Nakamura2018}, have been proposed as the confining medium responsible for jet collimation \citep{Nakamura2018, Chatterjee2019, Park2019a}. However, the black hole inflow-outflow system in 3C 84 may differ significantly from that of M87. Previous studies have suggested that the black hole in 3C 84 is also powered by hot accretion flows, given its luminosity being approximately 0.4\% of its Eddington luminosity \citep{Levinson1995, Plambeck2014}. Nevertheless, various observations provide evidence for the presence of substantial amounts of cold gas in the vicinity of the central engine in 3C 84 \citep{Vermeulen1994, Walker1994, Walker2000, FN2017, Kino2018, Kino2021, Wajima2020}. Our observations of the deflection of the jet knot and the presence of a dense ambient medium exhibiting FFA near the deflection point suggest a critical role for the cold gas within the central parsec in shaping the jet in 3C 84.

It is noteworthy to mention that different types of medium are responsible for the observed jet collimation at larger scales. At several mas scales, 3C 84 displays a nearly cylindrical profile in terms of jet width versus distance \citep{Nagai2014, Giovannini2018}. Recent RadioAstron space-VLBI observations at 5 GHz have revealed a cocoon-like structure surrounding the jet at the same scale, suggesting that a hot mini-cocoon could potentially serve as the ambient medium responsible for maintaining the cylindrical jet collimation profile observed at larger scales \citep{Savolainen2023}.

\subsection{Origin of the Abrupt Apparent Acceleration of the Knot}
\label{sec:discussion:acceleration}

In Section~\ref{sec:analysis:speed}, we demonstrated that the apparent speed of the knot experienced a sudden increase, transitioning from $\beta_{\rm app} = 0.32\pm0.01$ to $1.60\pm0.28$ during the mid-2022 period. Here we discuss possible origins of the observed acceleration. The observed apparent speed is given by: $\beta_{\rm app} = \beta\sin\theta_{\rm view} / (1 - \beta\cos\theta_{\rm view})$. We present the possible combinations of the two quantities ($\beta$ and $\theta_{\rm view}$) to reproduce the observed apparent speeds of $\beta_{\rm app} = 0.32$ and $1.60$ in the left panel of Figure~\ref{fig:speed}. Hence, the abrupt increase in the apparent speed could be attributed to an increase in the jet viewing angle, an increase in the intrinsic speed, or a combination of both factors.

The jet viewing angle of 3C 84 is uncertain, so is the intrinsic speed of the jet. Previous VLBI observations have suggested a relatively large jet viewing angle of $\theta_{\rm view} = 30-65^\circ$ based on the detection of the counterjet \citep{Walker1994, FN2017}. The counterjet was detected at relatively large distances of $\gtrsim2$ mas from the sub-parsec scale core \citep{Vermeulen1994, Walker1994, Walker2000, FN2017, Wajima2020}. On the other hand, modeling of the observed SEDs of 3C 84 have suggested a small jet viewing angle \citep{Abdo2009, TG2014, Aleksic2014} of $\theta_{\rm view} = 18-30^\circ$. Previous studies using KVN observations found that short-term variations in the $\gamma$-rays and radio bands are better correlated with the core region than the downstream \citep{Hodgson2018}, suggesting that the high-energy photons used in the SED modeling may originate from the jet at a much shorter distance than the scale where the counterjet was detected. A recent study \citep{Oh2022}, based on high-resolution multi-epoch observations of 3C 84 using the Global Millimeter VLBI Array (GMVA) at 86 GHz, has constrained the jet viewing angle to be less than $\sim35^\circ$ in the core region. Based on this result, it was suggested that the jet experiences a large bending between the core and the downstream region, reconciling the discrepancy in the jet viewing angles obtained in the previous studies. 

Interestingly, the observed deflection of the knot provides further support to this scenario. Our analysis reveals a substantial change in the position angle of the knot's trajectory, transitioning from $\sim-47^\circ$ to $\sim-9^\circ$ (Figure~\ref{fig:trajectory}). This finding suggests that the jet's viewing angle can indeed significantly vary between the core and the downstream jet region, consistent with the suggestion made in a previous study \citep{Oh2022}. However, it is worth noting that this scenario appears to require some fine-tuning. For instance, based on the left panel of Figure~\ref{fig:speed}, it necessitates an extremely small initial viewing angle, such as $\theta_{\rm view} \lesssim 4^\circ$, coupled with a high intrinsic jet speed of $\beta\gtrsim0.8$. This viewing angle is considerably smaller than the values of $\sim18-30^\circ$ favored by the SED modeling studies \citep{Abdo2009, Aleksic2014, TG2014}. Nonetheless, one of the studies found that a very small viewing angle of $6^\circ$ could fit the observed SED very well, although this model was not considered reliable due to the discrepancy with the larger viewing angles derived from the radio observations \citep{TG2014}.

If we entertain the notion that the jet viewing angle of 3C 84 is indeed very small near the core region, we can reconcile both the SED and the observed acceleration of the knot. In such a scenario, one might anticipate a sudden decrease in the knot's brightness due to the change in the Doppler factor after the abrupt acceleration. We illustrate the expected Doppler factor as a function of $\beta_{\rm app}$ and $\theta_{\rm view}$ in the right panel of Figure~\ref{fig:speed}. This indicates that the Doppler factor needs to slightly decrease from $\delta = 3-4$ to $\delta\lesssim3$ to explain the transition from $\beta_{\rm app} = 0.32$ with $\theta_{\rm view} \lesssim4^\circ$ to $\beta_{\rm app} = 1.60$ with $\theta_{\rm view} \gtrsim 20^\circ$. Notably, we find that the knot's intensity decreases faster with time after the acceleration (Figure~\ref{fig:intensity}), which appears consistent with this scenario, although the knot's brightness is also influenced by other factors, such as plasma heating and cooling within the knot, which are currently unknown, making it challenging to directly test the scenario.

An alternative scenario may be that the knot underwent some intrinsic acceleration during the deflection. This requires a large change in $\beta$, from $\beta\sim0.45$ to $\sim0.85$ for the smaller jet viewing angle range of $\theta_{\rm view} \approx 18-30^\circ$ and from $\beta\sim0.35$ to $\sim0.87$ for the larger jet viewing angle range of $\theta_{\rm view} \approx 30-65^\circ$. One of the most successful models that can explain the gradual jet accleration observed in other nearby AGN jets such as M87 \citep{Mertens2016, Park2019b}, Cygnus A \citep{Boccardi2016b}, and NGC 315 \citep{Park2021b} is the MHD jet acceleration mechanism \citep{Lyubarsky2009}. This mechanism predicts that jet collimation and acceleration occur simultaneously and jet acceleration occurs gradually at a distance $\lesssim10^4-10^6$ $R_g$ \citep{Marscher2008}. However, the observed acceleration of the knot was abrupt, and the 3C 84 jet shows a nearly cylindrical jet collimation profile \citep{Nagai2014, Giovannini2018}, unlike other AGN jets where the parabolic/semi-parabolic collimation profiles and gradual acceleration were observed to co-occur. These distinct characteristics suggest that a different mechanism may be required to explain the observed acceleration of the knot if it is indeed due to intrinsic acceleration. A recent study, based on jet kinematic analysis using the VLBA 43 GHz monitoring observations, has suggested that $\gamma$-rays observed in 3C 84 may be produced by magnetic-reconnection-induced mini-jets and turbulence \citep{Hodgson2021}. Magnetic reconnection can result in bulk acceleration of a blob within the jet as well as high-energy flares \citep{Giannios2009, Giannios2013}. It is interestig to note that 3C 84 started to show enhanced $\gamma$-ray activity since mid-2022 \citep{Fermi2023}, when the knot experienced acceleration, making this scenario one of the plausible explanations for the observed acceleration.

%% file: summary.tex
In this paper, we explore the intriguing radio source 3C 84, which experiences intermittent outbursts in radio flux density associated with the ejection of jet components from the sub-parsec scale core \citep{Nagai2010}. Recent monitoring observations with the VLBA at 43 GHz have revealed the ejection of a jet knot from the core in the late 2010s \citep{Punsly2021, Paraschos2022}. Intriguingly, this knot propagated in a direction significantly offset from the parsec-scale jet direction.

To investigate this phenomenon further, we present results from follow-up VLBA 43 GHz observations covering the period between October 2019 and November 2022. Employing the jet kinematic analysis method WISE \citep{ML2015}, which is suitable for studying complex jet structures \citep{Mertens2016, Park2019a, Hodgson2021}, we successfully traced the knot's trajectory during the monitoring period. We discovered that the knot underwent an abrupt change in its trajectory in the early 2020s, realigning itself with the parsec-scale jet direction.

Furthermore, we present results from a 22 GHz observation of 3C 84 with the GVA. By jointly analyzing the GVA 22 GHz image with a VLBA 43 GHz image observed about one week apart, we generated a spectral index map. Notably, the map reveals an inverted spectrum region near the edge of the jet where the knot experienced deflection. We attribute this inverted spectrum to FFA caused by a dense cold ambient medium characterized by an electron density of $n_{\rm amb} \sim 2.1\times10^5\ {\rm cm^{-3}}$. We have considered SSA as an alternative explanation but find it unlikely based on the brightness temperature of the jet near the knot deflection point and the inferred energy density ratio between synchrotron-emitting particles and the magnetic field. Notably, various observations have shown indications of FFA at different locations in the jet at parsec/sub-parsec scales \citep[e.g.,][]{Walker1994, Vermeulen1994, Walker2000, FN2017, Wajima2020, Kino2021}.

Our results highlight the crucial role of the surrounding ambient medium in shaping the jet. While previous studies have suggested non-relativistic gas outflows called winds as contributors to jet collimation in M87 and other AGNs \citep{Nakamura2018, Chatterjee2019, Park2019b, Park2021b}, our observations suggest that a dense cold medium may be responsible for the initial shaping of the 3C 84 jet, possibly associated with accretion flows, radiation-driven outflows, or a combination of both. It is noteworthy that the 3C 84 jet has been known to exhibit a nearly cylindrical jet collimation profile at larger scales \citep{Nagai2014, Giovannini2018}, and recent RadioAstron space-VLBI observations suggest strong collimation by a hot mini-cocoon surrounding the jet \citep{Giovannini2018, Savolainen2023}. Our results, combined with these observations, indicate that different types of ambient medium may influence the jet's shaping at different scales.

Furthermore, we observed a sudden increase in the apparent speed of the knot, from $\beta_{\rm app} = 0.32\pm0.01$ to $1.60\pm0.28$ during the mid-2022 period. We propose two plausible scenarios to explain this abrupt acceleration. One scenario involves a substantial change in the jet viewing angle between the core region and the downstream jet region, consistent with the significant deflection of the knot observed in our study, and the discrepancy in jet viewing angle constraints based on SED modeling \citep{Abdo2009, Aleksic2014, TG2014} and the detection of the counterjet \citep{Walker1994, FN2017}. However, this scenario may require fine-tuning of initial viewing angles and intrinsic jet speeds. The other scenario suggests intrinsic acceleration of the knot during the deflection, potentially driven by magnetic reconnection \citep{Giannios2009, Giannios2013}. Interestingly, 3C 84 displayed enhanced $\gamma$-ray activity since mid-2022 \citep{Fermi2023}, coinciding with the knot's acceleration, and magnetic reconnection is known to produce high-energy flares.

In conclusion, our study sheds light on the complex dynamics of the knot and its interaction with the ambient medium in shaping the 3C 84 jet. We emphasize the crucial role of the GVA, providing high angular resolution and excellent image fidelity through extensive collaboration among VLBI arrays worldwide. We advocate for more extensive utilization of the GVA in the future, which has the potential to yield groundbreaking results previously unattainable with a single VLBI array.

%% file: acknowledgements.tex
We express our appreciation to the referee for the constructive comments, which have significantly enhanced the quality of this paper. J.P. thanks Wajima Kiyoaki for useful comments and discussions. J.P. acknowledges ﬁnancial support through the EACOA Fellowship awarded by the East Asia Core Observatories Association, which consists of the Academia Sinica Institute of Astronomy and Astrophysics, the National Astronomical Observatory of Japan, Center for Astronomical Mega-Science, Chinese Academy of Sciences, and the Korea Astronomy and Space Science Institute. M.K. is support by the JSPS KAKENHI Grant number JP22H00157 and JP21H01137. H.N. is support by the JSPS KAKENHI Grant number JP21H01137 and JP18K03709. The European VLBI Network is a joint facility of independent European, African, Asian, and North American radio astronomy institutes. Scientific results from data presented in this publication are derived from the following EVN project code(s): GP060. This study makes use of VLBA data from the VLBA-BU Blazar Monitoring Program (BEAM-ME and VLBA-BU-BLAZAR; http://www.bu.edu/blazars/BEAM-ME.html), funded by NASA through the Fermi Guest Investigator Program. The VLBA is an instrument of the National Radio Astronomy Observatory. The National Radio Astronomy Observatory is a facility of the National Science Foundation operated by Associated Universities, Inc.

%% file: appendix.tex
\section{2D cross-correlation analysis}
\label{appendix:2dcc}

In Section~\ref{sec:analysis:kinematics}, we made the assumption that the western component of the nuclear structure acts as a stationary core. To further validate this assumption, we conducted a two-dimensional cross-correlation analysis \citep{CG2008} on the downstream jet emission located at a declination less than $-1.5$ mas between each pair of successive epochs. We considered two cases: first, assuming the stationary core to be the western component in the core region ("Ref 1"), and second, considering the stationary core as the eastern component identified by WISE ("Ref 2"). For instance, in the case labeled as "Ref 1," the images from all epochs are aligned in a manner that positions the core at the map's origin, mirroring the approach undertaken for producing the movies (Figure~\ref{fig:movie}). Conversely, in the "Ref 2" scenario, the images are aligned so that the knot resides at the map's origin. In both instances, we determined the displacement between images of consecutive epochs that maximizes the cross-correlation coefficient.

\begin{figure}[t!]
\centering
\includegraphics[width=\linewidth]{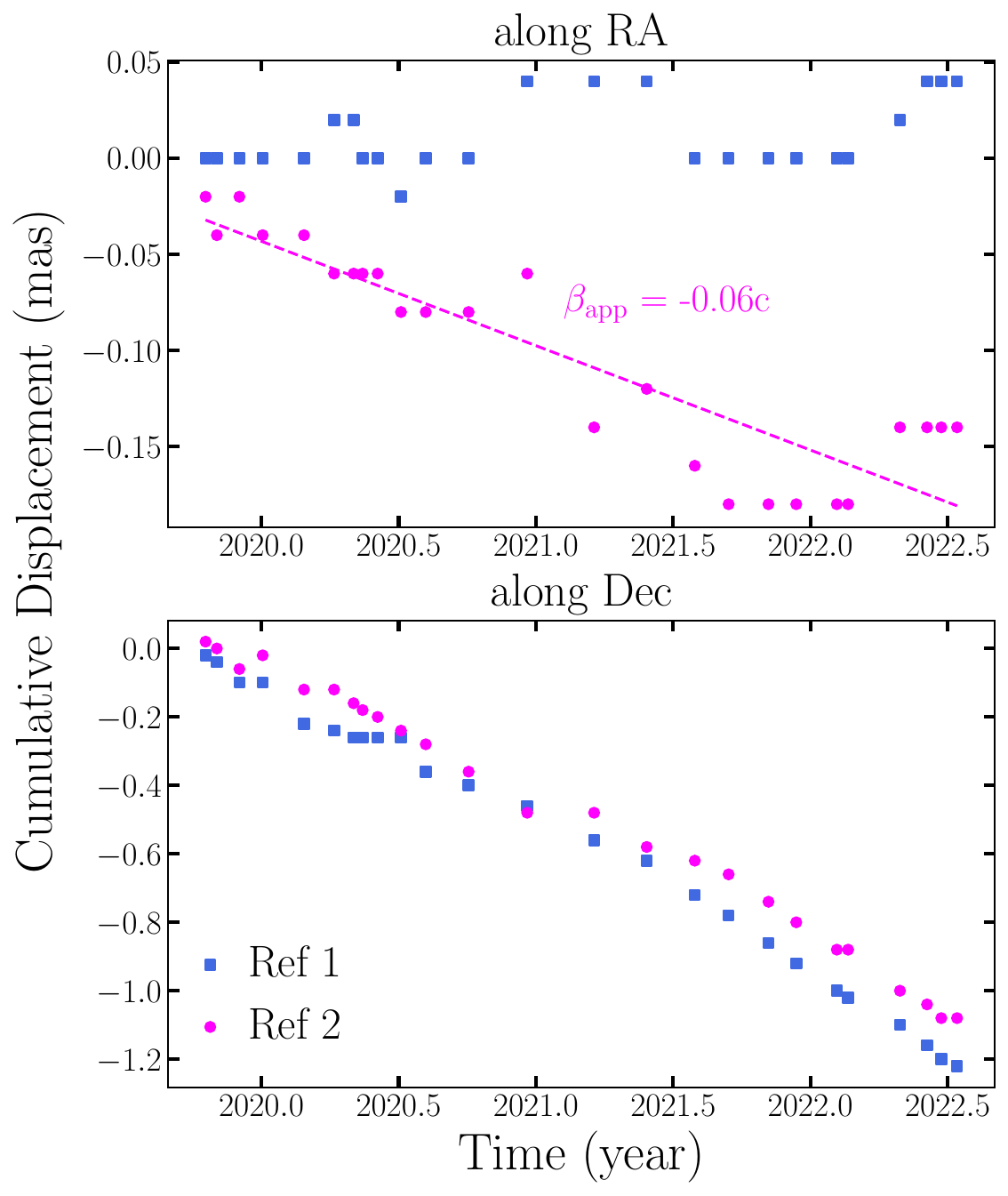}
\caption{Cumulative displacement as a function of time, maximizing the cross-correlation coefficients of the downstream jet emission located at a declination less than -1.5 mas between each pair of successive epochs, shown for the right ascension direction (top) and declination direction (bottom). The blue data points corresponds to the result obtained by assuming the western component in the core region as the stationary core ("Ref 1"), while the magenta data points corresponds to the result obtained by assuming the eastern component, identified through WISE analysis, as the stationary core ("Ref 2").}
\label{fig:reference}
\end{figure}

Figure~\ref{fig:reference} illustrates the cumulative displacement as a function of time. In both cases, the cumulative displacement gradually decreases over time along the declination direction, potentially indicating a gradual downward motion of the downstream jet, which will be described in detail in an upcoming paper (Kam et al., in prep.). However, there is a notable distinction between the two cases along the right ascension direction: while the cumulative displacement for "Ref 2" gradually decreases over time, the cumulative displacement for "Ref 1" remains near zero. This result suggests that, in the case of "Ref 2", the downstream jet collectively moves westward, perpendicular to the jet direction. We attribute this behavior to the selection of an incorrect reference point, and we conclude that "Ref 1", which served as the reference point in our kinematic analysis, is the correct choice.

\section{Color maps of the GVA 22 GHz and VLBA 43 GHz images}
\label{appendix:gva}

\begin{figure}[t!]
\centering
\includegraphics[width=\linewidth]{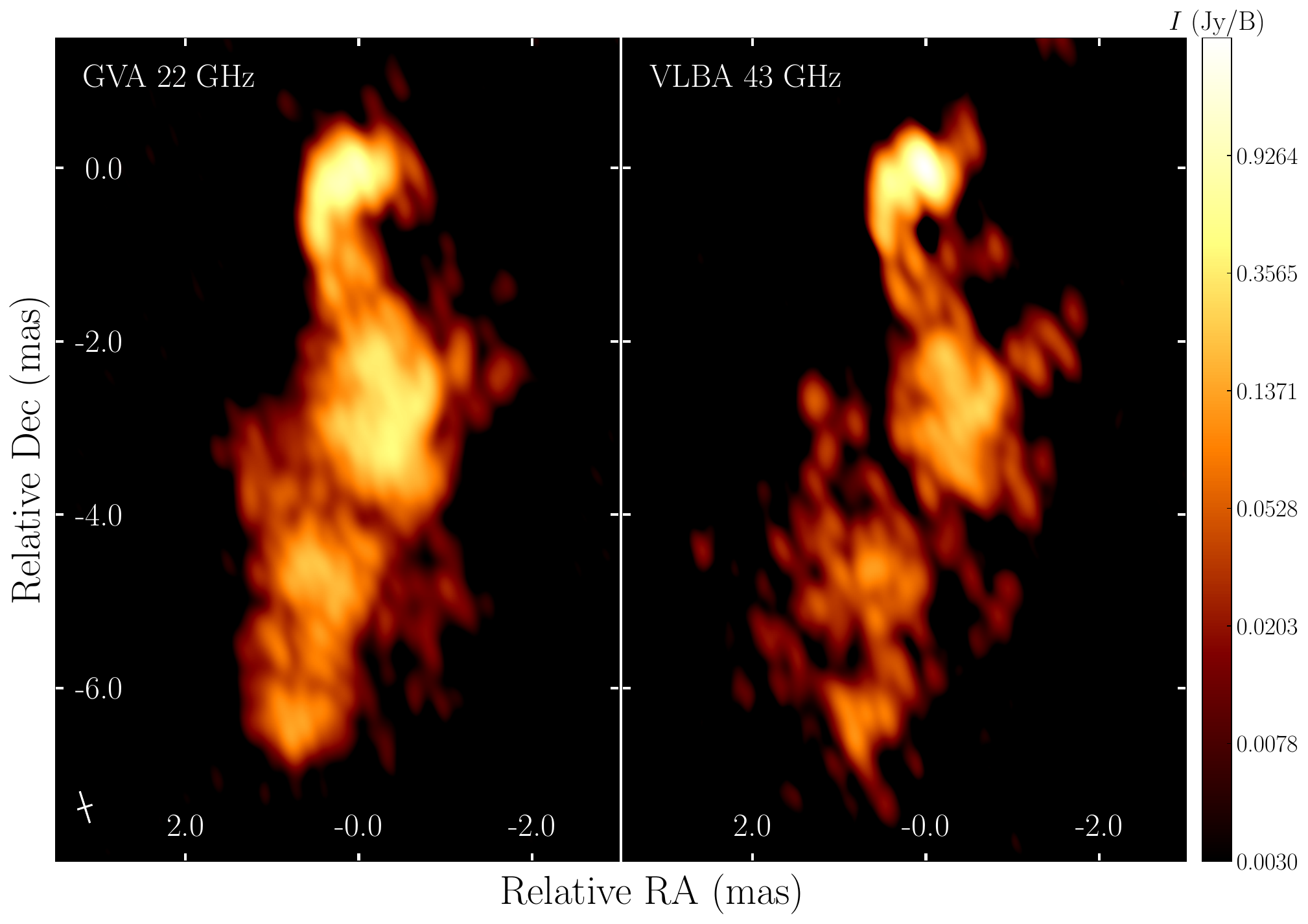}
\caption{Color maps of the GVA 22 GHz (left) and VLBA 43 GHz (right) images employed to generate the spectral index maps in Figures~\ref{fig:spix} and ~\ref{fig:spix_entire}. The size of the restoring beam is indicated by a white cross in the lower-left corner of the left image.}
\label{fig:gva}
\end{figure}

In Figure~\ref{fig:gva}, we present color maps of the GVA 22 GHz and VLBA 43 GHz images, which were utilized in generating the spectral index map detailed in Section~\ref{sec:analysis:spix}.

\section{Spectral index error map}
\label{appendix:spixerr}

\begin{figure}[t!]
\centering
\includegraphics[width=\linewidth]{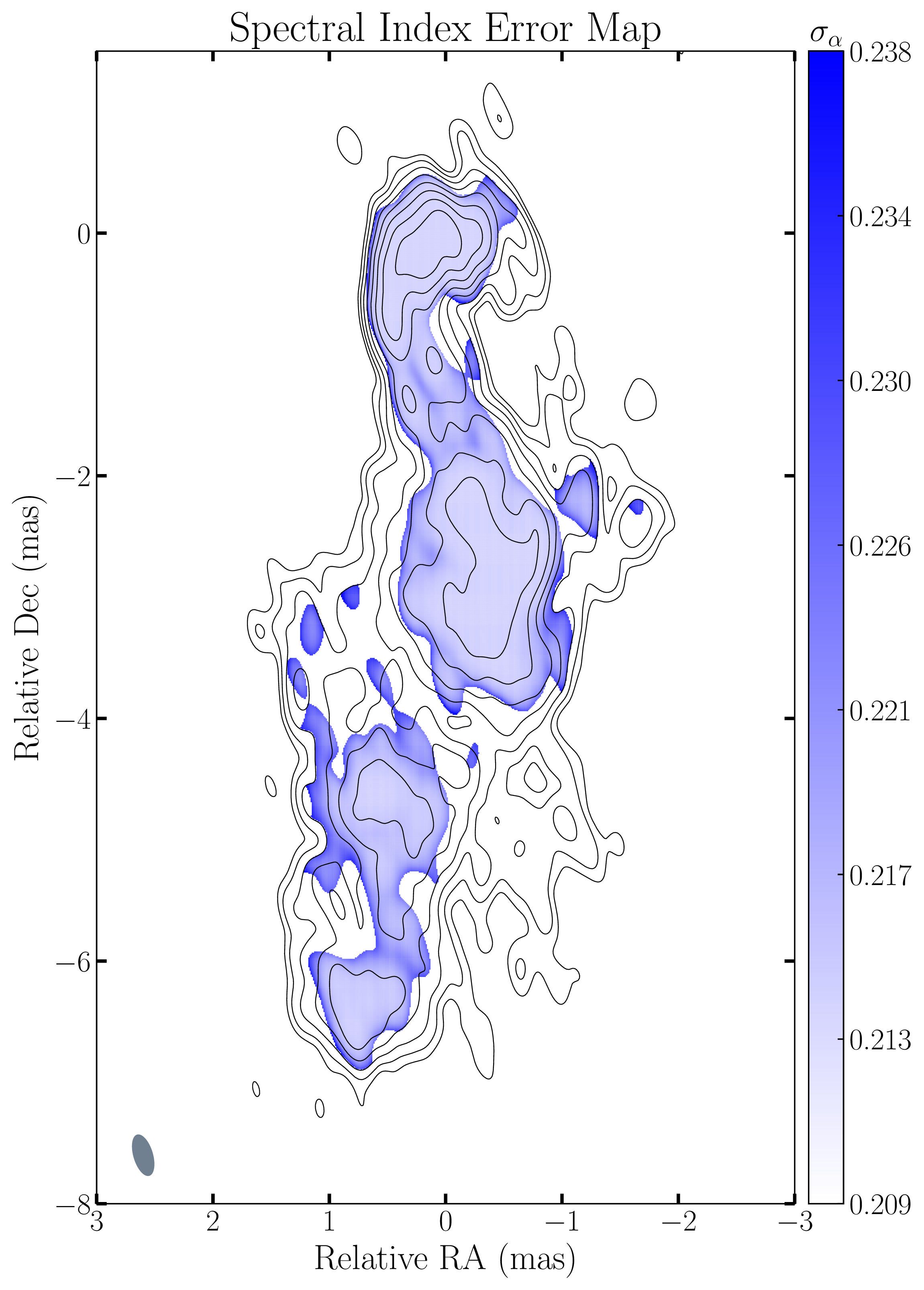}
\caption{Distribution of the spectral index error, acquired by combining the errors attributed to the image rms noise \citep{KT2014} and the systematic errors arising from the uncertainties in visibility amplitudes, through quadrature.}
\label{fig:spixerr}
\end{figure}

In Figure~\ref{fig:spixerr}, we present the distribution of the spectral index errors. These errors are not notably greater than the systematic errors, with $\sigma_{\alpha, {\rm sys}}\approx0.21$, which stem from the presumed 10\% systematic amplitude errors. This observation is due to our computation of spectral index values for pixels with intensities exceeding 20 times the off-source image rms noise, as outlined in Section~\ref{sec:analysis:spix}.

\section{Stability of the inverted spectrum in relation to restoring beam size}
\label{appendix:beam}

\begin{figure*}[t!]
\centering
\includegraphics[width=0.24\linewidth]{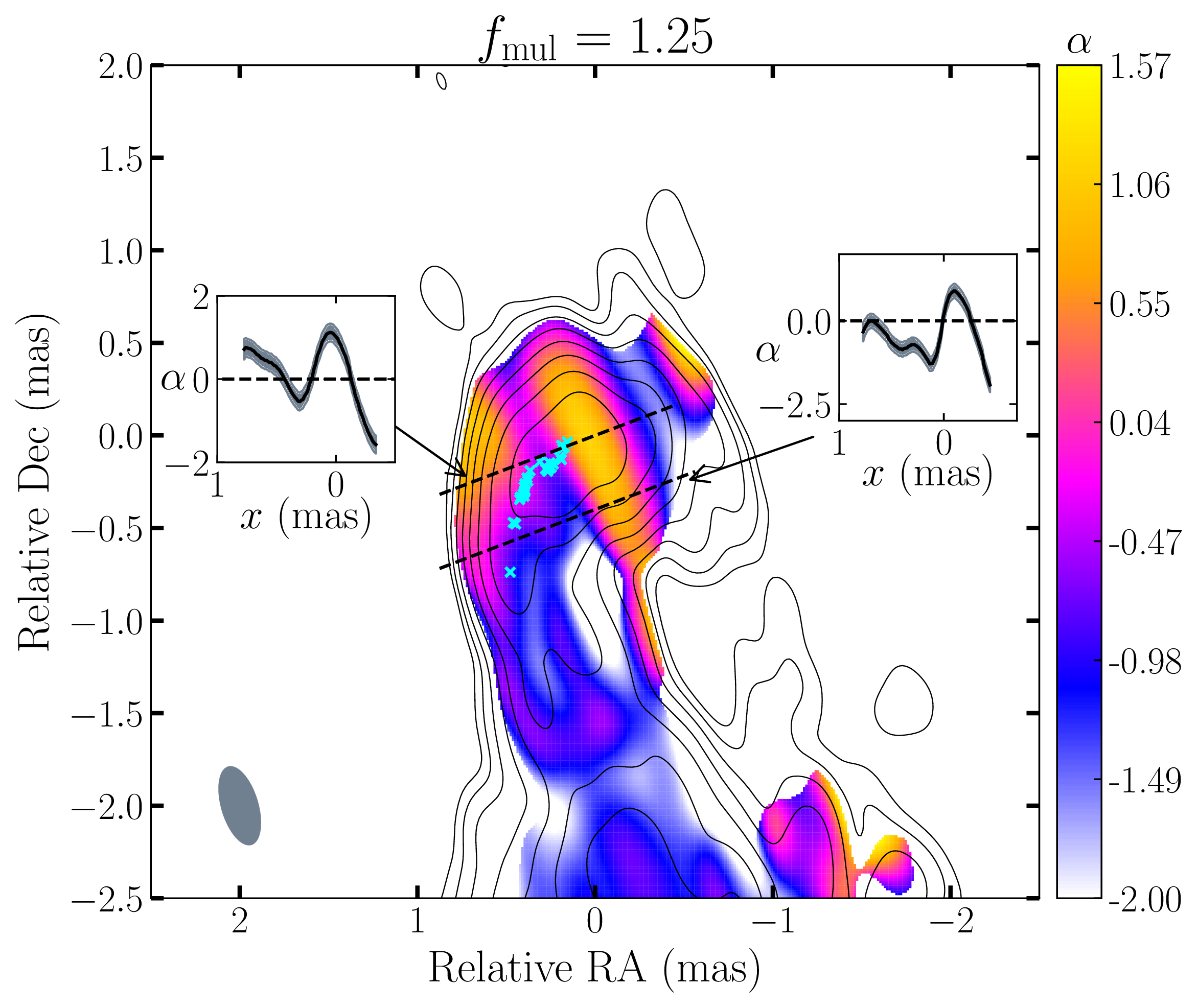}
\includegraphics[width=0.24\linewidth]{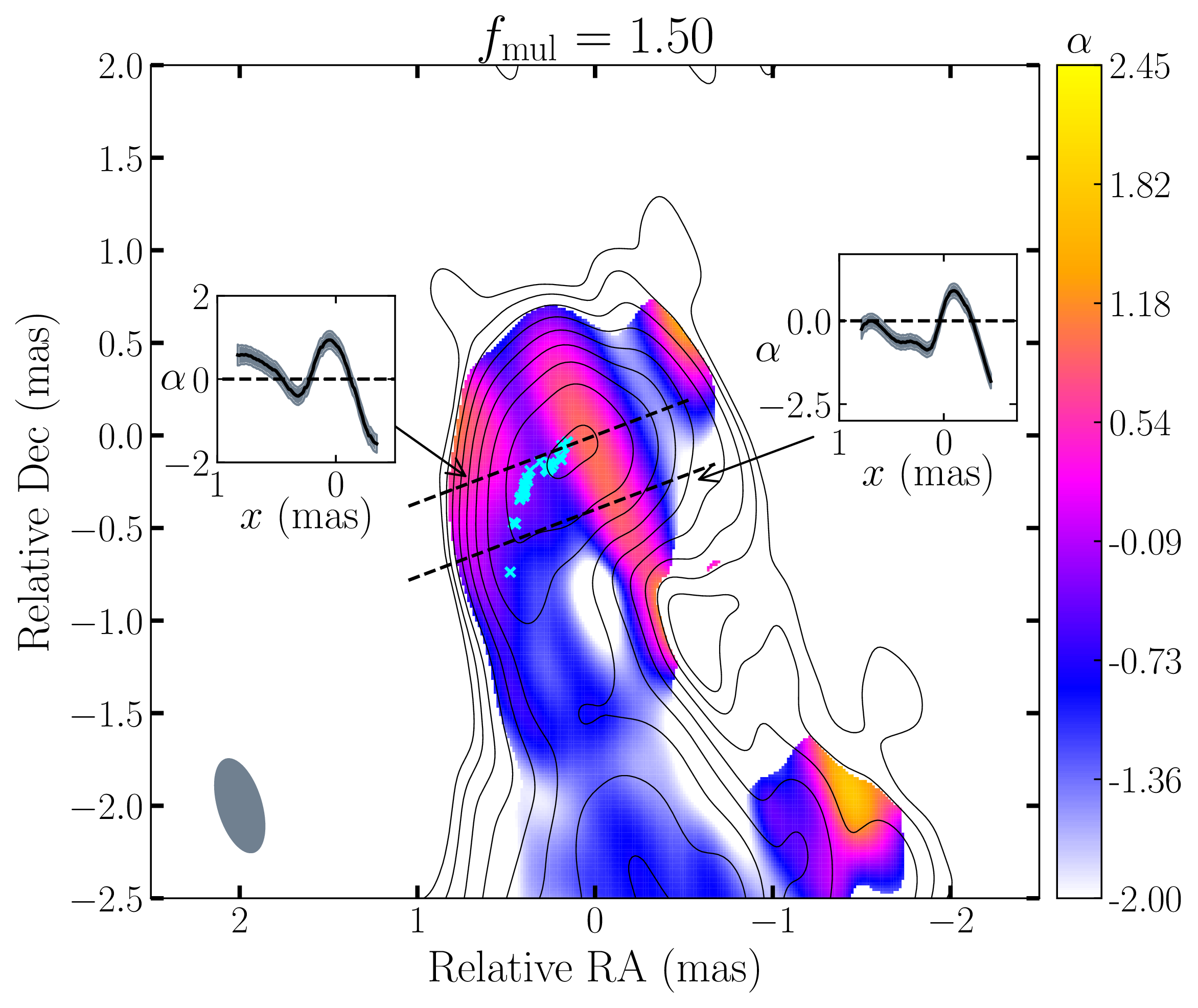}
\includegraphics[width=0.24\linewidth]{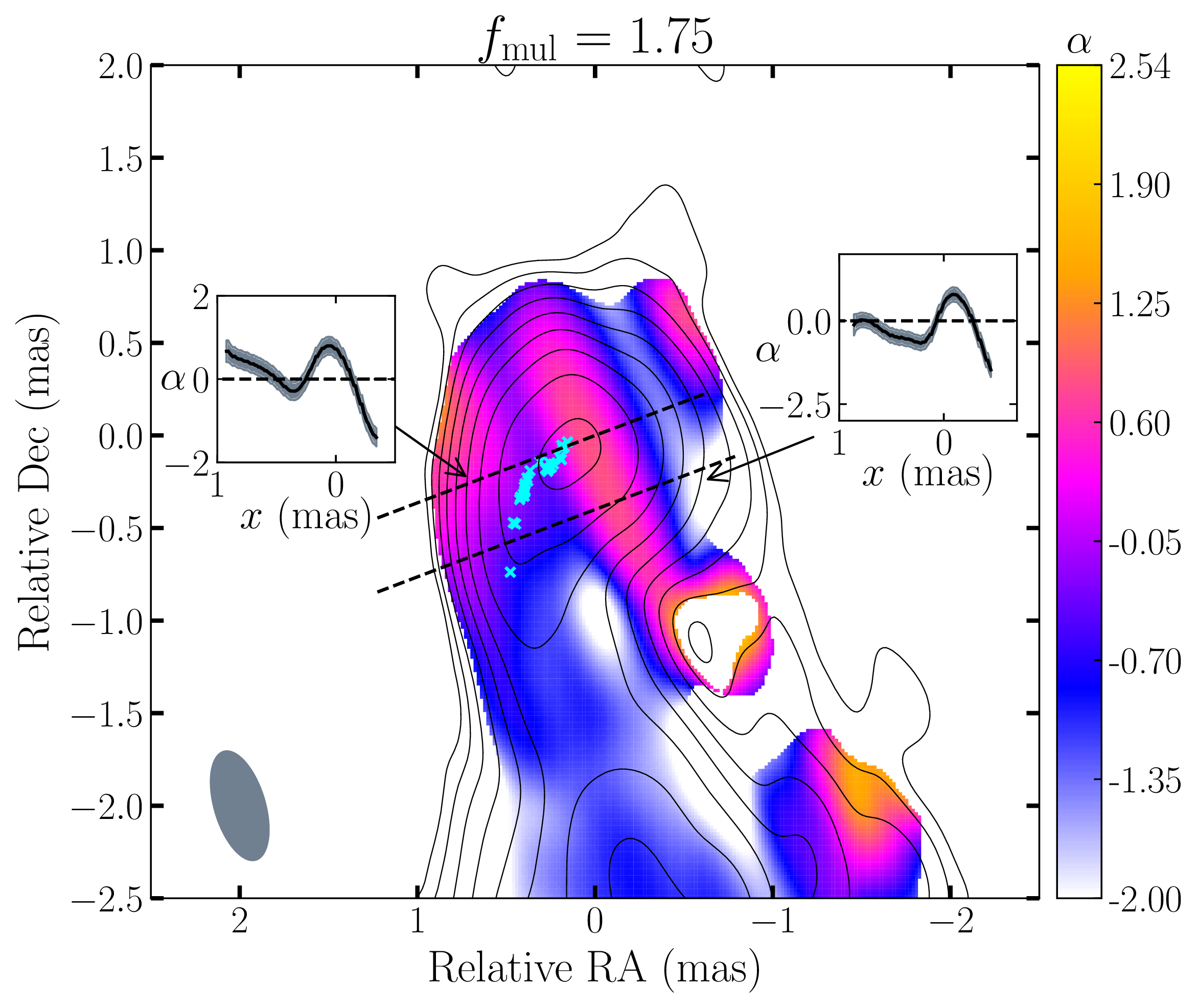}
\includegraphics[width=0.24\linewidth]{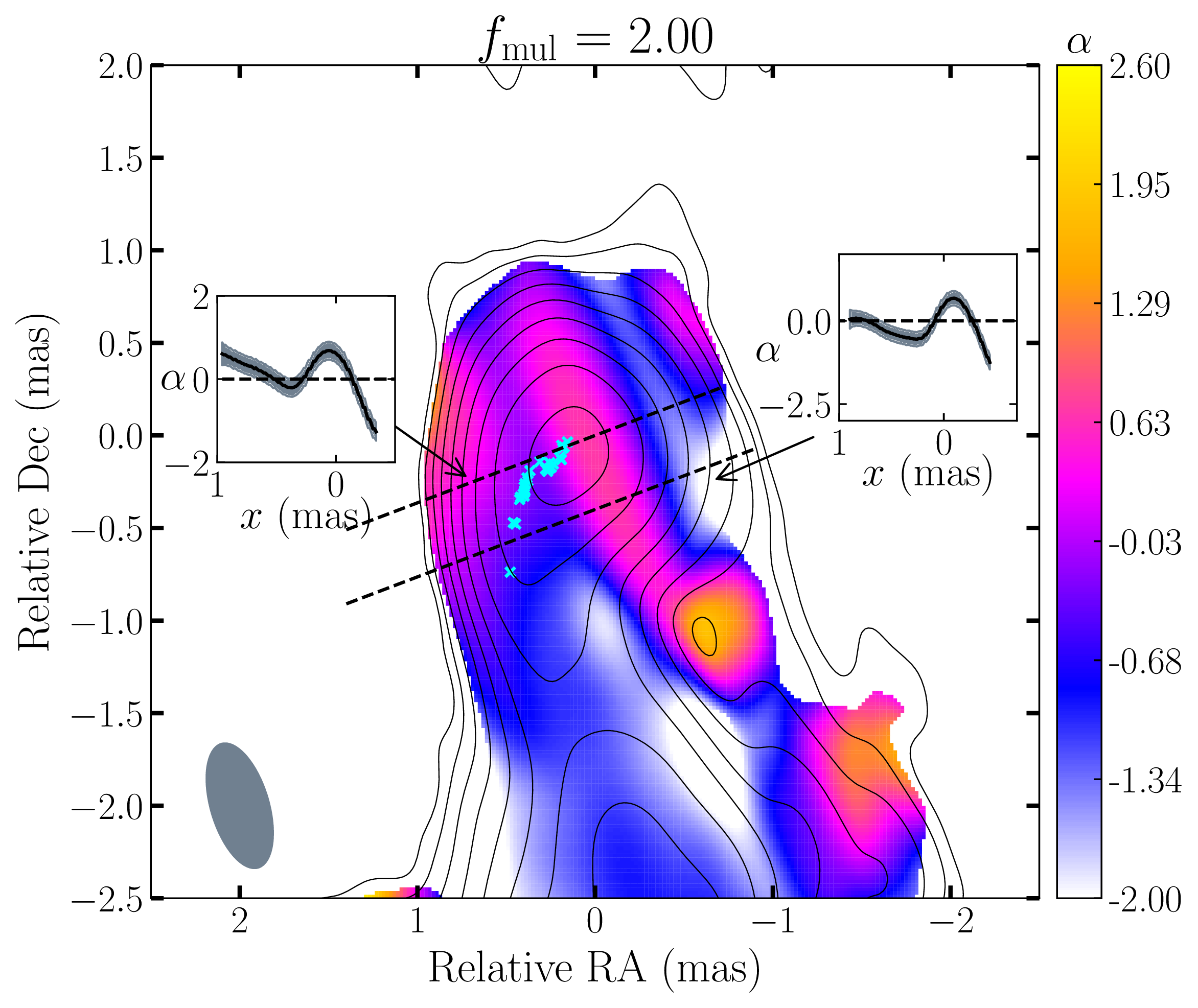}
\caption{Spectral index maps generated by applying the identical analysis as conducted in Section~\ref{sec:analysis:spix}, but using the 22 and 43 GHz images with a restoring beam size equal to the original beam size multiplied by $f_{\rm mul}$. The results for $f_{\rm mul} = 1.25, 1.50, 1.75, 2.00$ are displayed from left to right. The position of the knot identified through WISE analysis is indicated by cyan crosses. The presence of the inverted spectral region near the knot's deflection point persists even with a larger restoring beam size, although the contrast diminishes in comparison to the original spectral index map (Figure~\ref{fig:spix}).}
\label{fig:spix_beam}
\end{figure*}

In this Appendix, we test if the inverted spectral region near the knot deflection point (Figure~\ref{fig:spix}) persists if we increase the restoring beam size. We derived the spectral index map by performing the same analysis as done in Section~\ref{sec:analysis:spix} using the 22 and 43 GHz images, employing a restoring beam size of original beam size multiplied by $f_{\rm mul}$, where $f_{\rm mul}$ is a multiplication factor. We explored four cases with $f_{\rm mul}$ values of 1.25, 1.50, 1.75, and 2.00, and present the resulting images in Figure~\ref{fig:spix_beam}. We found that the inverted spectrum region remains present even for larger beam sizes in general, but as expected, the contrast becomes weaker.

\section{Stability of the inverted spectrum against potential image misalignment}
\label{appendix:misalignment}

In this appendix, we investigate the persistence of the inverted spectral region near the knot's deflection point (see Figure~\ref{fig:spix}) when we intentionally introduce artificial misalignment between the 22 and 43 GHz images. The primary goal of this analysis is to confirm the robustness of the conclusions presented in our paper with respect to potential image alignment artifacts resulting from imperfect alignment procedures.

\begin{figure*}[t!]
\centering
\includegraphics[width=0.24\linewidth]{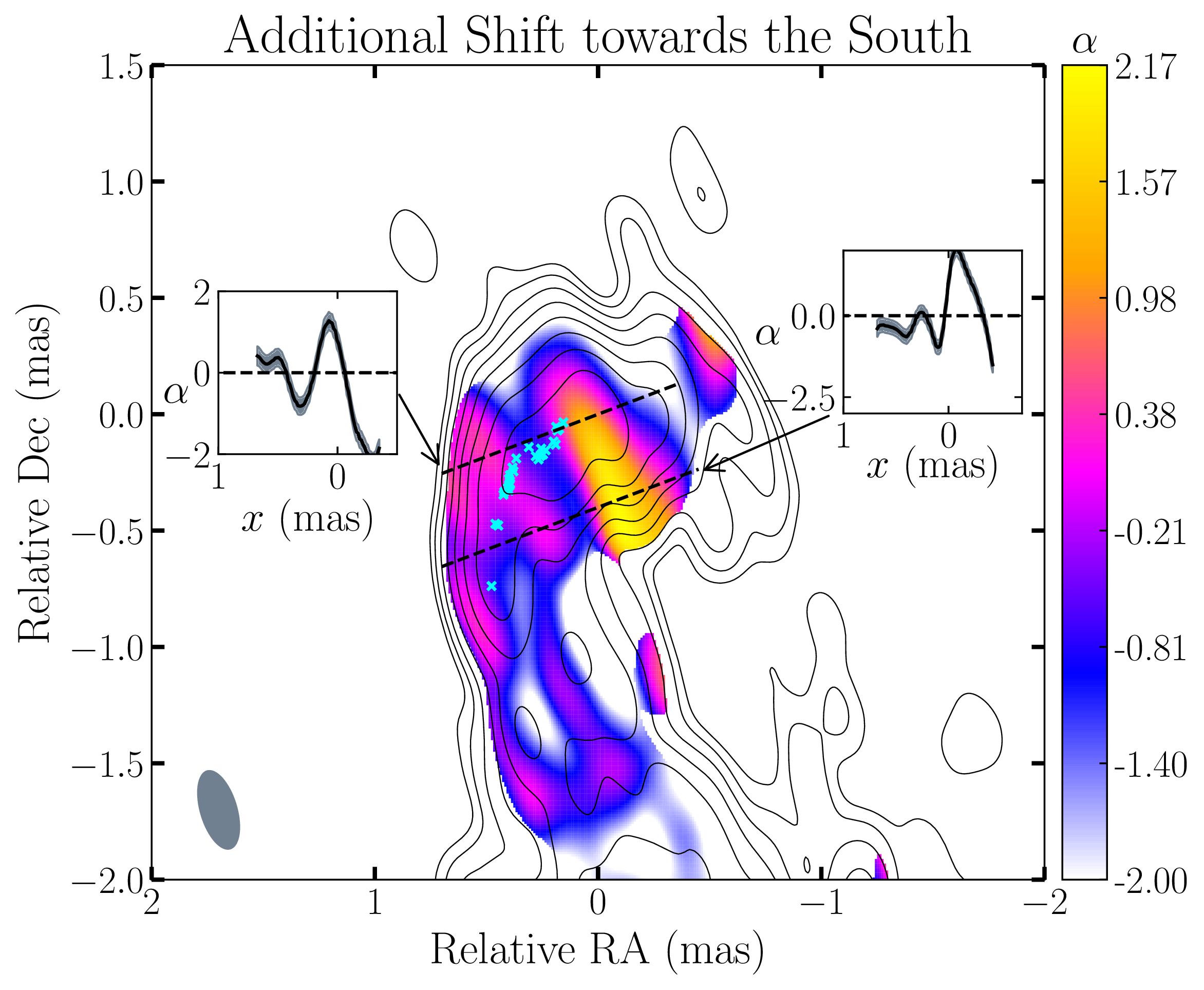}
\includegraphics[width=0.24\linewidth]{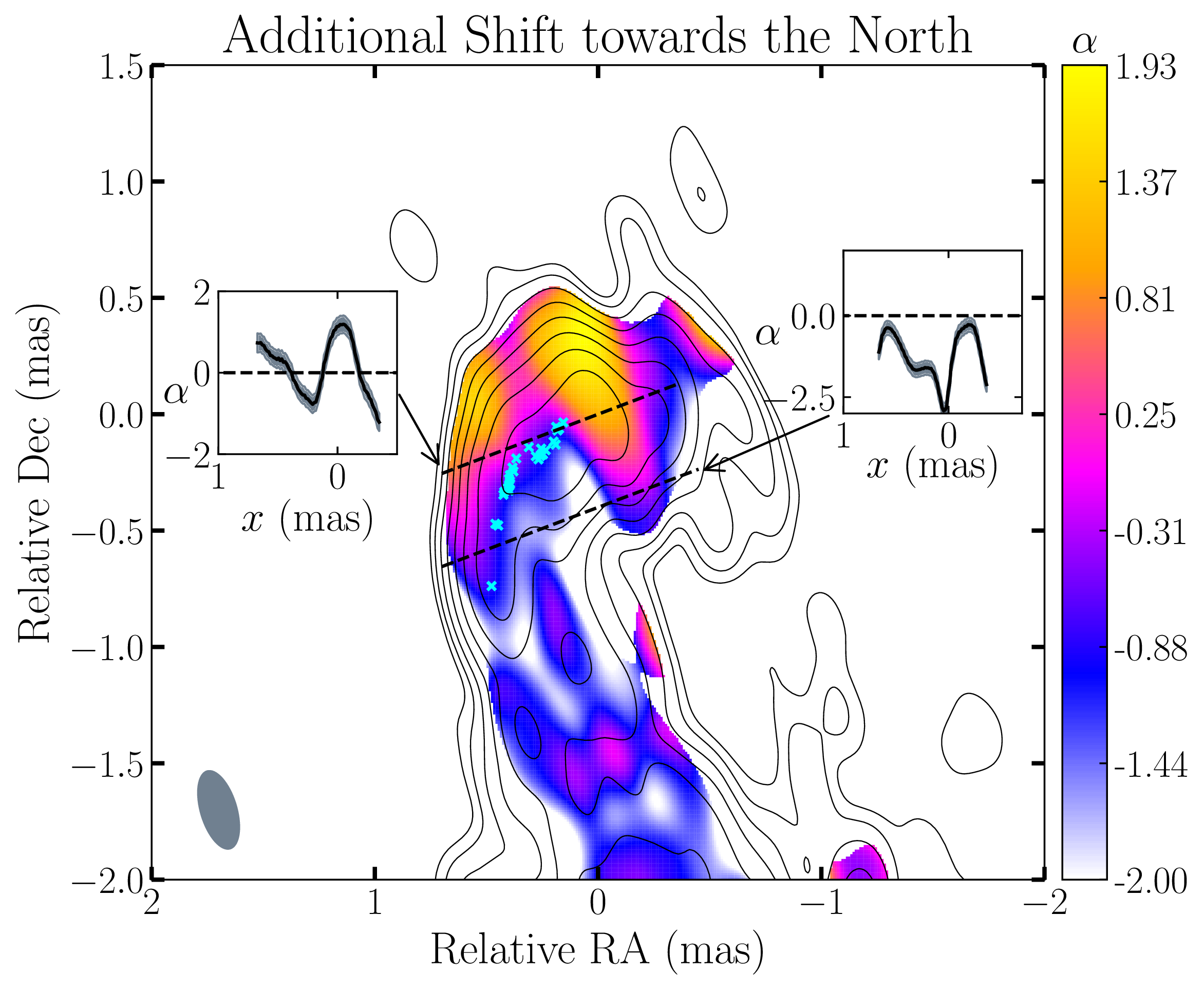}
\includegraphics[width=0.24\linewidth]{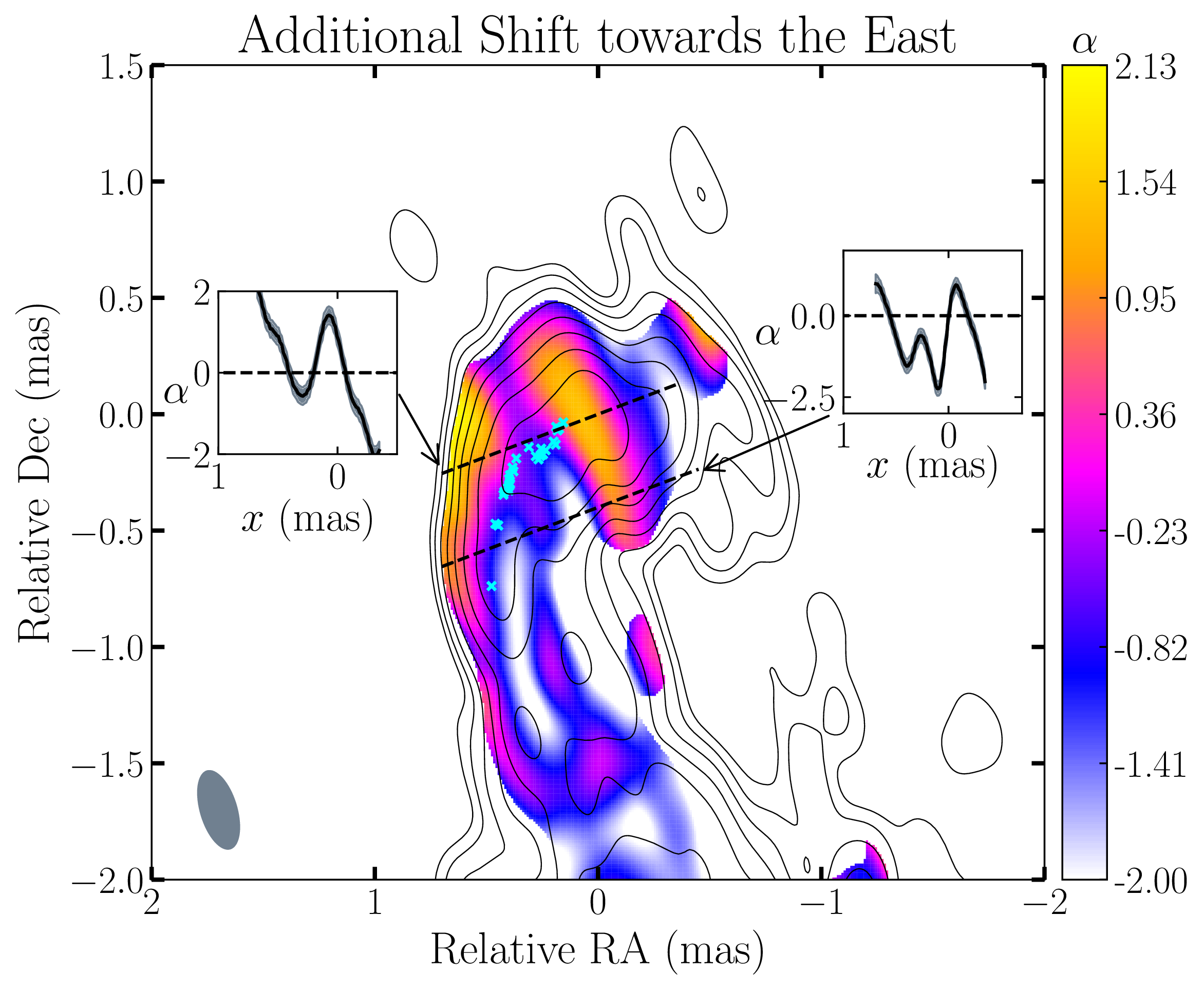}
\includegraphics[width=0.24\linewidth]{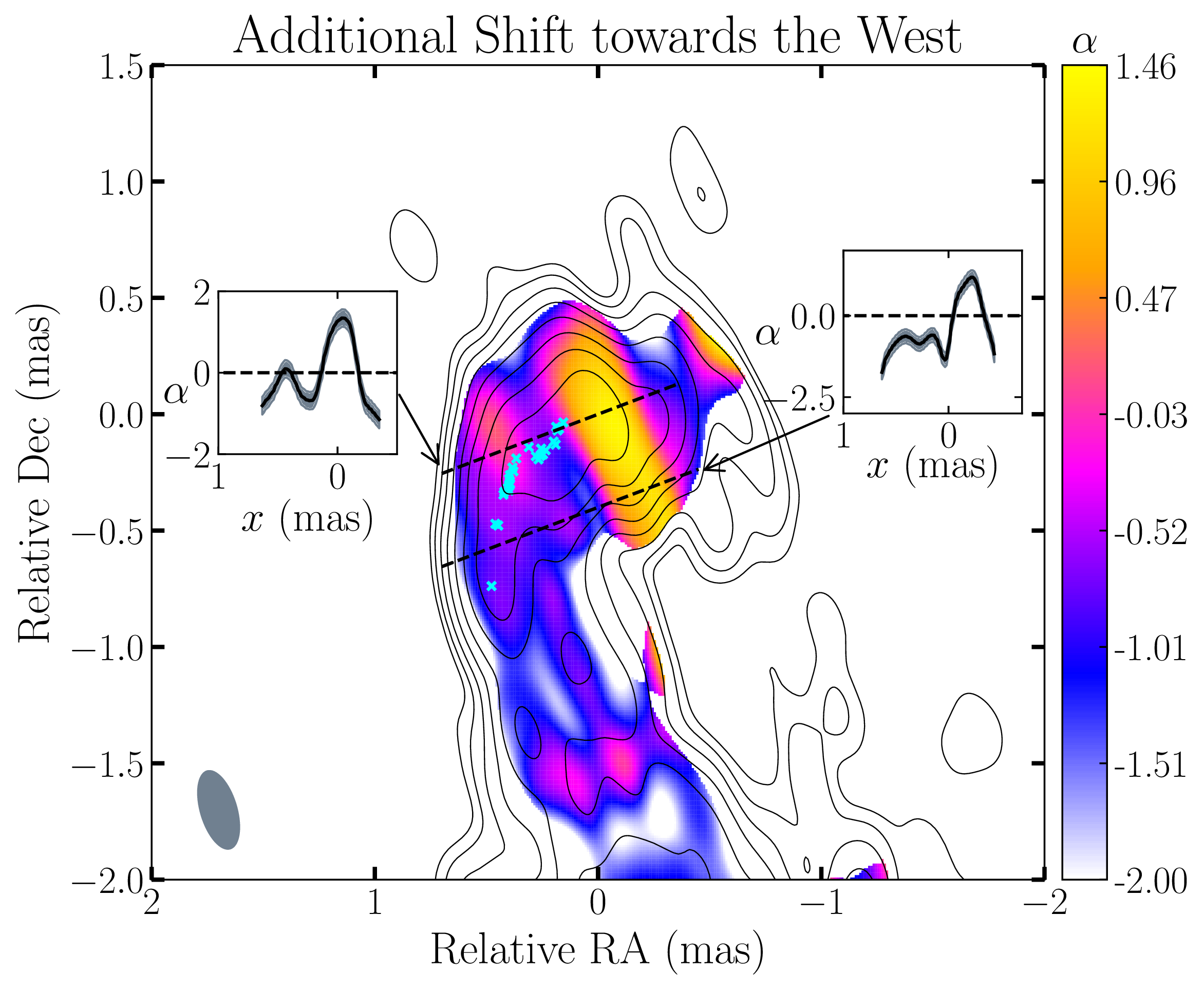}
\caption{Spectral index maps were generated using the same analysis techniques as detailed in Section~\ref{sec:analysis:spix}. However, in this analysis, we introduced an additional shift in the 43 GHz image in multiple directions, including South, North, East, and West (from left to right). Each shift was approximately 1/4 of the synthesized beam size in its respective direction. Cyan crosses indicate the position of the knot identified through WISE analysis. Notably, the presence of the inverted spectral region near the knot's deflection point persists even when there is a substantial amount of image misalignment. However, it's worth mentioning that, in some cases, the contrast is somewhat reduced compared to the original spectral index map (see Figure~\ref{fig:spix})..}
\label{fig:spix_misalignment}
\end{figure*}

We generated the spectral index map using the same methodology as outlined in Section~\ref{sec:analysis:spix}. However, in this analysis, we applied an additional shift to the 43 GHz image in the South, North, East, and West directions. The magnitude of the shift was approximately 1/4 of the synthesized beam size in the respective direction. The resulting images are presented in Figure~\ref{fig:spix_misalignment}. Our findings demonstrate that the inverted spectral region near the knot's deflection point remains visible even in the presence of significant image misalignment. However, it is important to note that in some cases, the contrast is somewhat reduced compared to the original spectral index map.

%\section{\jp{Investigating the potential influence of limited $(u,v)$-coverage on detecting the inverted spectral region through imaging simulations}}
\section{Investigating whether the inverted spectral region is an artifact}
\label{appendix:synthetic}

While the GVA 22 GHz data and the VLBA 43 GHz data employed in deriving the spectral index map (Section~\ref{sec:analysis:spix}) share similar synthesized beam shapes, their $(u,v)$-coverages exhibit distinctions. This variance could potentially lead to artifacts within the images, particularly in regions characterized by low SNRs. In this appendix, we delve into the possibility of the inverted spectral region near the knot's deflection point arising due to differences in $(u,v)$-coverage through an imaging simulation.

We employed GPCAL \citep{Park2021a} to generate synthetic data. In this appendix, we provide a concise explanation of the synthetic data generation procedure (see \citealt{Park2021c, Park2023a, Park2023b} for more details). The complex visibilities observed between two VLBI antennas, denoted as $m$ and $n$, are formulated within a visibility matrix, $\V_{mn}$, depicted as:
\begin{equation}
\V_{mn} = 
\begin{pmatrix}
r^{RR}_{mn} & r^{RL}_{mn}\\
r^{LR}_{mn} & r^{LL}_{mn}
\end{pmatrix}.
\end{equation}
Here, $R$ and $L$ denote right- and left-handed circular polarizations (RCP and LCP), respectively. The observed $V_{mn}$ experiences corruption due to multiple factors, and it is convenient to arrange of all these corruptions into a unified Jones matrix \citep{Jones1941}. For the sake of simplicity, we make the assumption that the Jones matrix coincides with the antenna gain matrix. As a result, other matrices like the polarization leakage matrix become inconsequential. This assumption remains valid, considering our primary emphasis on the analysis of total intensity data. In this context, the Jones matrix takes on the following form:
\begin{eqnarray}
    \J_{m} = \G_m =
    \begin{pmatrix}
    G^R_m & 0 \\
    0 & G^L_m
    \end{pmatrix},
\end{eqnarray}
where $G$ is the complex antenna gain. Subscripts and superscripts denote antenna numbers and polarization, respectively. 

The observed $\V_{mn}$ values represent modifications of the true $\bar\V_{mn}$ through:
\begin{equation}
\label{eq:rime}
    \V_{mn} = \J_m \bar\V_{mn} \J^H_n,
\end{equation}
where H represents the Hermitian operator. In the case of circular feeds, the $\V$ values are linked to the Fourier transforms of the Stokes parameters ($\tilde{I}$, $\tilde{Q}$, $\tilde{U}$, and $\tilde{V}$) through:
\begin{equation}
\label{eq:stokes}
\bar\V_{mn} \equiv
\begin{pmatrix}
\mathscr{RR} & \mathscr{RL} \\
\mathscr{LR} & \mathscr{LL}
\end{pmatrix}
=
\begin{pmatrix}
\tilde{I}_{mn} + \tilde{V}_{mn} & \tilde{Q}_{mn} + j\tilde{U}_{mn} \\
\tilde{Q}_{mn} - j\tilde{U}_{mn} & \tilde{I}_{mn} - \tilde{V}_{mn}
\end{pmatrix}.
\end{equation}
For the sake of simplicity, we assume $\tilde{Q} = \tilde{U} = \tilde{V} = 0$. Additionally, we consider the Stokes $I$ CLEAN model from the GVA 22 GHz data as the ground-truth source model. This implies that the Fourier Transform of the CLEAN model corresponds to $\tilde{I}$ in Equation~\ref{eq:stokes}. To generate synthetic data, we created a dataset with the same $(u,v)$-coverage as the actual GVA 22 GHz and VLBA 43 GHz data (Section~\ref{sec:analysis:spix}). Accounting for the uncertainties associated with each real data point, we introduced thermal noise to the corresponding synthetic data points.

We conducted tests under two scenarios: one assumed unity antenna gains ($G = 1$), while the other considered non-unity gains ($G \neq 1$). The former assessed the influence of different $(u,v)$-coverages between 22 and 43 GHz, as well as the potential imperfections in the CLEAN imaging process. The latter extended the investigation to include the effects of antenna gain corruptions. We assumed that antenna gains fluctuate from scan to scan while remaining constant within a scan. Gain amplitudes for each station were randomly drawn from a Gaussian distribution with a standard deviation of 0.1. Additionally, an overall shift in gain amplitudes away from unity was introduced, with each station undergoing a uniform random adjustment between $-0.2$ and $0.2$. In line with the synthetic data generation methodology employed in EHT analyses \citep[e.g.,][]{EHT2019d}, we incorporated a random station phase for each scan to emulate the absence of an absolute phase.

In the case of the former, we performed imaging using Difmap, subsequently aligning the images through 2D cross-correlation. This process was followed by the creation of a spectral index map, mirroring the procedure outlined in Section~\ref{sec:analysis:spix}. In the context of the latter, we extended our approach by incorporating iterative imaging and self-calibration through Difmap.

\begin{figure*}[t!]
\centering
\includegraphics[width=0.48\linewidth]{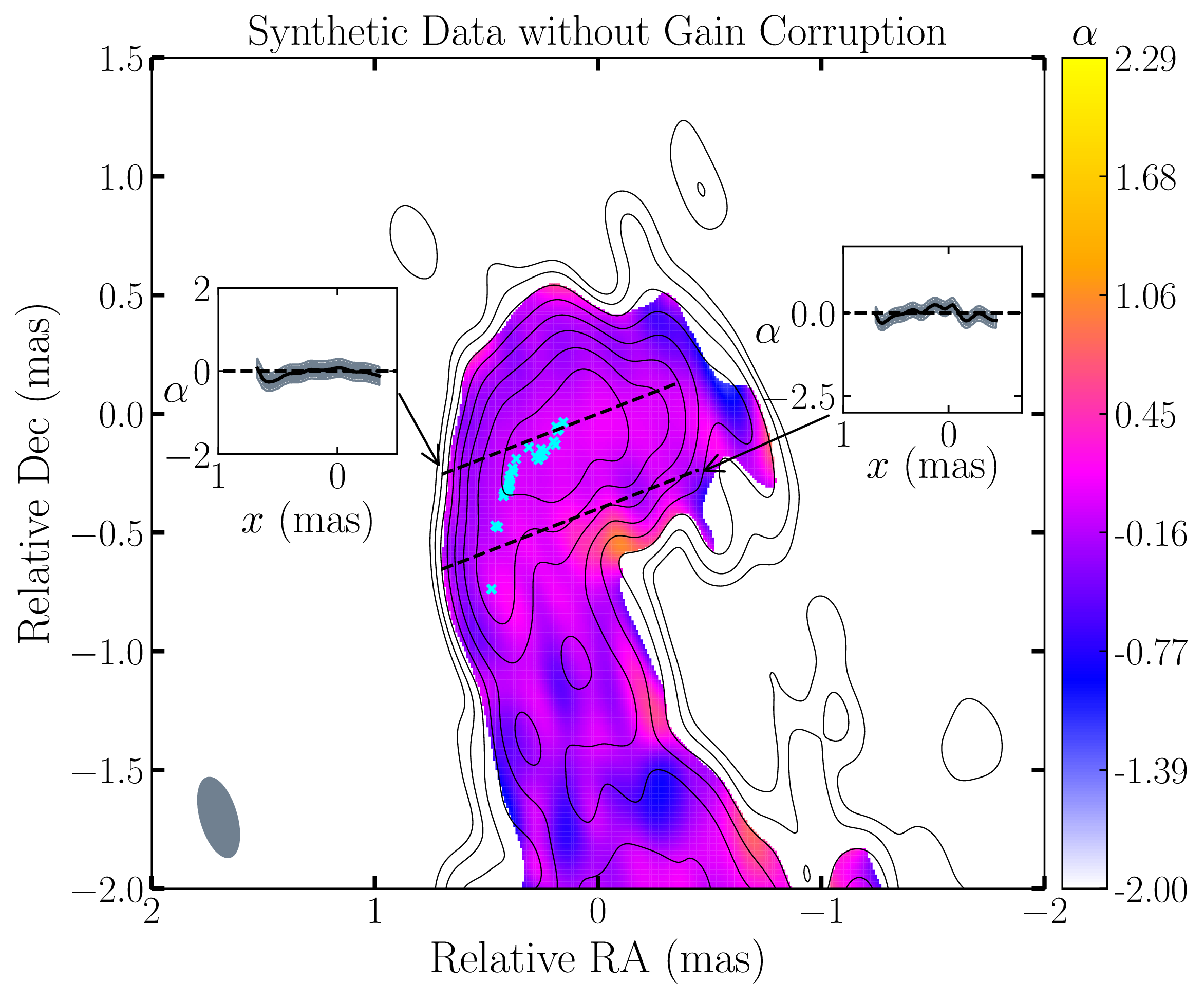}
\includegraphics[width=0.48\linewidth]{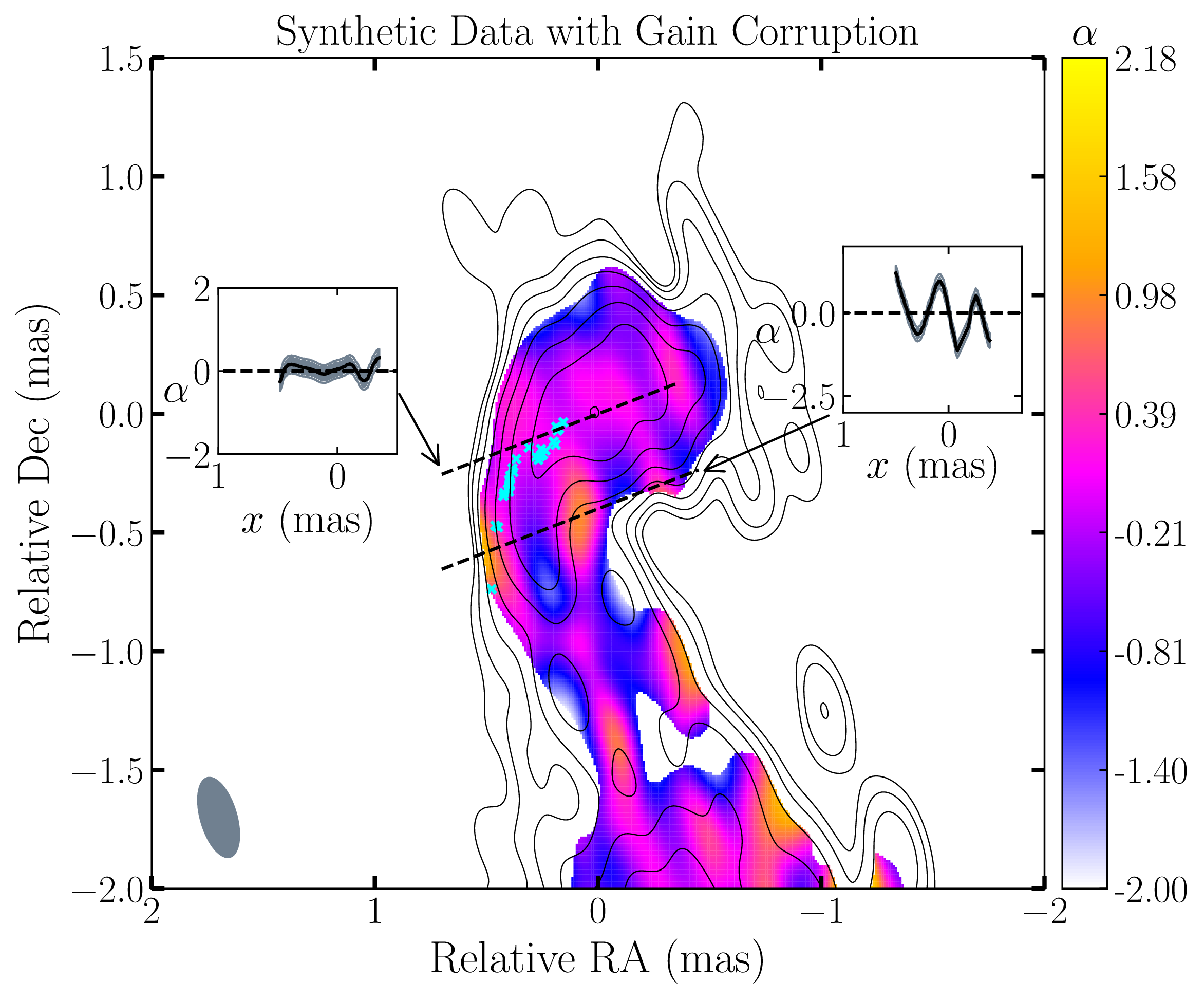}
\caption{Spectral index maps generated by applying the identical analysis as conducted in Section~\ref{sec:analysis:spix}, but using synthetic data without (left) and with (right) gain corruptions. Deviation from a zero spectral index can be ascribed to different $(u,v)$-coverages between the 22 and 43 GHz data, along with imperfections in gain calibration and imaging procedures. While deviations are indeed evident in certain regions of the jet, particularly in the image on the right due to gain corruptions, there is no indication of a significantly inverted spectrum near the knot's deflection point, as observed in the real data.}
\label{fig:synthetic}
\end{figure*}

The resultant spectral index maps are depicted in Figure~\ref{fig:synthetic}. Given that identical ground-truth models were employed for generating both the 22 and 43 GHz synthetic datasets, any deviation from a zero spectral index can be attributed to imperfections within the synthetic data as well as potential limitations in the CLEAN imaging and self-calibration procedures. When considering synthetic data without gain corruptions, the spectral indices exhibit a general distribution around zero, indicating that the difference in $(u,v)$-coverages between 22 and 43 GHz itself has a minor influence on the resulting spectral index map. Conversely, for synthetic data incorporating gain corruptions, the distortion becomes more pronounced, indicating that the effects of imperfect gain calibration can contribute additional uncertainties to the spectral index map. However, it is important to note that in both cases, no significantly inverted spectral region arises near the knot's deflection point. This implies that the observed inverted spectral region is not an artifact; rather, it signifies a genuine manifestation of an FFA absorbing medium, which is accountable for the knot's deflection.

